\documentclass[prd,aps,showpacs,groupedaddress,amsmath,amssymb,nofootinbib,superscriptaddress,floatfix,noeprint,10pt]{revtex4-2}
\usepackage{silence}
\usepackage{mathrsfs}
\usepackage{amsmath}
\usepackage{amsfonts}
\usepackage{amssymb}
\usepackage{array}
\usepackage{verbatim}
\usepackage{epsfig}
\usepackage{graphicx} 
\usepackage[usenames, dvipsnames]{color}
\usepackage{marginnote}
\usepackage{slashed}
\usepackage{bm}
\usepackage{todonotes}
\usepackage{tikz}
\usepackage{dsfont}
\usepackage{braket}

\usepackage{todonotes}
\usepackage{xcolor}
\usepackage{placeins}
\usepackage[colorlinks=true,allcolors=blue,bookmarks=false]{hyperref}
\usepackage{orcidlink}
\usepackage{lineno}    % line number
\begin{document}

\title{{Thermodynamic characteristics of a Fermi gas with an invariant energy scale and its astrophysical implications}}

%--------------------------------------------------------------------------
\author{Tiyasa~Kar \orcidlink{0000-0002-6476-1226}}
\email{tkar@ncsu.edu}
\affiliation{Department of Physics and Astronomy, North Carolina State University, Raleigh, NC 27695, USA.}

%--------------------------------------------------------------------------
\author{Atul~Kedia \orcidlink{0000-0002-3023-0371}}
\email{akedia@nccu.edu}
\affiliation{Department of Physics and Astronomy, North Carolina State University, Raleigh, NC 27695, USA.}
\affiliation{Department of Mathematics and Physics, North Carolina Central University, Durham, NC 27707, USA.}
%--------------------------------------------------------------------------
\author{Ramkumar~Radhakrishnan \orcidlink{0000-0001-6838-9153}}
\email{rradhak2@ncsu.edu}
\affiliation{Department of Physics and Astronomy, North Carolina State University, Raleigh, NC 27695, USA.}
%--------------------------------------------------------------------------
%%%%%%%%%%%%%%%%%%%%%%%%%%%%%%%%%%%%%%%%%%%%%%%%%%%%%%%%%%%%%%%%%%%%%%%%%%%
\begin{abstract}
We investigate the thermodynamics of a relativistic Fermi gas governed by a modified dispersion relation in the Magueijo Smolin (MS) formulation of Doubly Special Relativity (DSR), characterized by the presence of an invariant ultraviolet energy (deformation) scale. We study the system in two physically distinct regimes: the near degenerate low-temperature limit, and the high-temperature regime. In the low-temperature regime, we derive the thermodynamic quantities using the standard Sommerfeld expansion. In the high-temperature regime, we evaluate all thermodynamic quantities numerically from the exact grand canonical potential and demonstrate that the thermodynamics of the Fermi gas reduces to the standard relativistic ideal gas behavior. We apply the resulting low-temperature equation of state to study compact astrophysical objects, namely, non-rotating white dwarfs and neutron stars. Helium white dwarfs exhibit a strong dependence on the deformation scale, while white dwarfs composed of heavier elements are less affected. For neutron stars, the modified equation of state leads to configurations that are smaller in radius and lower in mass than that is produced by nucleonic equations of state. Our results highlight how modified relativity theories can be probed by studying astrophysical objects.

\end{abstract}
%%%%%%%%%%%%%%%%%%%%%%%%%%%%%%%%%%%%%%%%%%%%%%%%%%%
% \date{\today}
\maketitle
 %\tableofcontents
%
\section{Introduction} \label{sec:intro}
Physical phenomena in nature occur over a vast range of length and energy scales, and almost all fundamental theories reflect this hierarchy. Such theories are therefore best understood as effective theories \cite{kaplan2005five}, valid only up to a certain cutoff scale. By isolating a particular regime—where parameters associated with smaller scales vanish and those corresponding to larger scales diverge—effective field theory provides an overall and a parametric framework for describing physical systems within well defined limits. 

Within this perspective, quantum gravity is naturally expected to demonstrate itself as an effective theory characterized by a fundamental scale. Any formulation of quantum gravity requires the existence of either a smallest (but finite) length scale ($l$) or an upper bound on energy ($\kappa$). Since relativity demands that such fundamental scales be observer independent, their introduction directly conflicts with Einstein’s special relativity \cite{Einstein1921-EINRTS-2}, where Lorentz transformations make lengths and energies to be observer dependent. If different inertial observers were to measure different fundamental scales, the notion of a universal effective theory would break down, violating the equivalence principle \cite{amelino1998tests}. Consequently, a deformation of special relativity becomes necessary.

One prominent realization of this idea is Doubly Special Relativity (DSR), originally proposed by Amelino-Camelia \cite{amelino1998tests}. DSR extends special relativity by introducing, in addition to the invariant speed of light $c$, a second observer independent scale typically an ultraviolet energy cutoff $\kappa$. Since its establishment, several formulations and simplified models of DSR have been developed (see refs. \cite{ghosh2010recent, kowalski2005introduction} for reviews and refs. \cite{Magueijo_2002, magueijo2003generalized, amelino2002relativity,rosati2015planck,husain2016low,lobo2017rainbows,girelli2007modified,bojowald2010quantum} for further discussions). A universal feature of these approaches is the emergence of a modified dispersion relation (MDR) \cite{Magueijo_2002,magueijo2003generalized}, describing deviations from standard relativistic kinematics while preserving observer independence. The existence of an ultraviolet cutoff is also strongly motivated by independent developments in black hole physics, string theory, loop quantum gravity, and non commutative geometry \cite{maggiore1993generalized, park2008generalized, snyder1947quantized, gross1988string, kempf1997minimal, kober2010minimal,verma2019effect,radhakrishnan2022wigner}. Among various realizations of DSR, the formulation introduced by Magueijo and Smolin (MS) \cite{Magueijo_2002} is particularly appealing: while the Lorentz algebra itself remains unchanged, its representation on phase space becomes nonlinear, naturally incorporating the invariant energy scale $\kappa$. Standard special relativity is recovered smoothly in the limit $\kappa \rightarrow \infty$. Recent refinements and applications of this framework further emphasize its relevance in high energy and gravitational contexts \cite{2024AnPhy.47069826K,mishra2018invariant}. 

In this work, we adopt the MS formalism and investigate the thermodynamics of matter governed by a modified dispersion relation with an explicit ultraviolet energy cutoff. Unlike conventional approaches where deviations appear only asymptotically at very high energies, the MS model induces corrections that can persist even at comparatively low energies. The thermodynamics of classical and bosonic systems—such as ideal gases and photon gases—within this framework has been studied extensively (see refs. \cite{chandra2012thermodynamics,das2024ideal,zhang2011photon,camacho2007thermodynamics,alexander2004noncommutativegeometryrealizationvarying, PhysRevD.81.025005,GREGG_2009}). Moreover, the astrophysical and cosmological implications of MDR-corrected thermodynamics have been explored in a variety of contexts \cite{amelino2004severe,amelino2012uv,camacho2006white,alexander2004noncommutativegeometryrealizationvarying,PhysRevD.81.025005}. In contrast, the thermodynamics of fermionic systems within the DSR framework has received comparatively less attention, despite its direct relevance to compact astrophysical objects. Degenerate Fermi gases provide the microscopic foundation of white dwarfs and neutron stars, making their thermodynamic properties of fundamental importance. Previous studies have primarily focused on the strictly degenerate $T = 0$ limit \cite{Mishra_2017}. While mathematically convenient, this limit represents an idealization that is never realized in physical systems. In astrophysical environments, compact stars possess a small but finite temperature, with $T<<T_{F}$, where $T_{F}$ is the Fermi temperature. The near-degenerate limit $T\rightarrow 0$ therefore offers a more physical description, allowing controlled thermal corrections while retaining the essential features of degeneracy pressure. We will investigate the thermodynamics of a Fermi gas governed by the MS modified dispersion relation in two physically distinct regimes: the near-degenerate limit $T\rightarrow 0$ and the high- temperature regime. The former is directly relevant for cold astrophysical objects such as white dwarfs and neutron stars, while the latter extends the applicability of the MDR framework to hot and dense environments. Using a method distinct from earlier treatments, we derive the corresponding thermodynamic quantities and explicitly recover the standard special relativistic results in the appropriate limits. Therefore, the relatively small values of $\kappa$ employed in this study should be interpreted as an effective phenomenological parameter rather than a universal quantum-gravity scale.

Corrections arising from modified dispersion relations can alter the equilibrium structure of compact objects, leading to observable deviations in their macroscopic properties \cite{2021PhLB..82236684S,2024AnPhy.47069826K}. Employing the MDR-corrected thermodynamics derived in this work, we analyze the structure of white dwarfs and neutron stars and compute the corresponding mass–radius relations, highlighting regimes where departures from standard special relativistic predictions may become significant. Degenerate matter in white dwarf and neutron star is susceptible to substantial modifications in the dispersion relation as it determines the thermodynamic relationships of nucleonic matter. 

The article is organized as follows: In Section~\ref{Sec:Moddispersion}, we introduce the modified dispersion relation arising within the DSR framework and outline its essential features relevant to our work. In Section~\ref{sec:thermo}, we derive the thermodynamic properties of a Fermi gas obeying this modified kinematics in two distinct regimes: the near degenerate limit ($T\rightarrow 0$) and the high-temperature regime. The results from near degenerate limit are then employed in Section~\ref{sec:astrophysics} to investigate their astrophysical implications, particularly on the structure and mass-radius relations for white dwarfs and neutron stars. Finally, Section~\ref{sec:conclusion} summarizes our main findings and presents concluding remarks.

\section{Aspects of modified dispersion relation}\label{Sec:Moddispersion}
The usual dispersion relation \cite{Einstein1921-EINRTS-2} for a particle in Special relativity is given by
\begin{align}
    \label{eq:SR}
    E^{2} = p^{2} + m^{2},
\end{align}
where $E$ is the energy, $p$ is the three-momentum and $m$ is the rest mass. Throughout this work we adopt natural units, $c = K_{B} = \hbar = 1$. In order to work within the class of DSR theories, one must reformulate the usual dispersion relation (Eq. \eqref{eq:SR}) using the deformed algebra characterized by an observer independent invariant energy scale $\kappa$. Therefore we specify the corresponding modified dispersion relation, which in the MS basis takes the form
\begin{align}
\label{eq:modified}
E^{2} & = p^{2} + m^{2}\left(1 - \frac{E}{\kappa}\right)^{2},
\end{align}
where $E$ and $p$ denote the particle energy and momentum, respectively, and $\kappa$ has the dimensions of energy. This reduces to the standard dispersion relation (Eq. \eqref{eq:SR}) when $\kappa\rightarrow\infty$.For finite $\kappa$, the relation implies that, for a fixed momentum, the particle energy is reduced relative to its special relativistic value, since
\begin{align}
    0\leq \bigg(1-\frac{E}{\kappa}\bigg)^{2}\leq 1.
\end{align}
It is important to explicitly distinguish between the fundamental DSR deformation scale and the effective deformation scale employed in the present work. In the standard Doubly Special Relativity (DSR), the deformation parameter $\kappa$ is expected to be a universal, observer independent scale associated with quantum-gravity effects, and is typically identified with the Planck energy, $\kappa \sim 10^{19}\, \mathrm{GeV}$. Although current observational constraints from high energy astrophysical phenomena—such as $\gamma$-ray bursts, ultra-high-energy cosmic rays and other astroparticle measurementsm have not detected deviations from Lorentz invariance, they place lower bounds on the deformation scale at $\kappa~\gtrsim~10^{9}\,~\mathrm{GeV}$\,\cite{amelino2013quantum,vasileiou2013constraints}. The scale considered in our present work should not be interpreted as this fundamental Planckian DSR scale. Rather, it is an effective in medium parameter that characterizes the collective behavior of matter under the extreme density and pressure conditions found in compact stars. The parameter $m$ appearing in Eq.~\eqref{eq:modified} is interpreted as an invariant mass as it remains unchanged under DSR transformations. This differs from special relativity, where $m$ corresponds directly to the rest mass. The physical rest mass energy $m_{0}$ is obtained by setting $p = 0$ in Eq.~\eqref{eq:modified}, yielding
\begin{equation} \label{eq:restmass}
    m_{0} = \frac{m}{1+\frac{m}{\kappa}}.
\end{equation}

As given in Ref.~\cite{magueijo2003generalized}, physical states in the MS framework are restricted to energies $E<\kappa$. Within this domain, the requirement that the rest mass be non-negative ($m_{0}\geq 0$) implies that the invariant mass is bounded as $0\leq m < \infty$. Consequently, the physically admissible region of the modified dispersion relation is given by
\[
0 \leq p < \kappa, \qquad
0 \leq E < \kappa, \qquad
0 \leq m < \infty .
\]
In the limits $\kappa\rightarrow \infty$ and $m\rightarrow 0$, the MS dispersion relation reproduces the standard special relativistic and massless cases, respectively. In the following section, we use this modified kinematical framework as the basis for constructing the thermodynamics of a Fermi gas in the near degenerate ($T\rightarrow 0$) and high-temperature regimes.

\section{Thermodynamic quantities of Fermi gas }\label{sec:thermo}
In this section we discuss the thermodynamic properties of a relativistic Fermi gas within the modified dispersion framework. Our goal is to obtain explicit expressions for the pressure, entropy, internal energy, and specific heat in the relevant physical limits. Since the behavior of the system is controlled by parameters such as temperature $(T)$, particle mass $(m)$, and ultraviolet deformation scale $(\kappa)$, we consider separately the low- temperature (nearly degenerate) regime and the high- temperature regime. In each case we adopt the appropriate statistical ensemble and perform controlled expansions to extract the leading thermodynamic behavior.
\subsection{At nearly degenerate limit ($T\rightarrow 0$)}\label{sec:T=0}
We make use of the standard Sommerfeld expansion in order to arrive at the expressions for thermodynamic quantities at very low but finite temperature. At this temperature range, the step function is no longer clean but is smeared around the Fermi energy.  This method of expansion takes care of it by resulting in the emergence of correction terms, in the orders of temperature, to the degenerate case. The dependence of these quantities on the number density of the particles arises from the chemical potential which changes with temperature and is no longer equal to the Fermi energy, but some deviation from it. So, the first term in this expansion of the thermodynamic parameters corresponds to the degenerate case and the subsequent terms are corrections to them. We will be considering corrections up to the order of $T^2$.
\subsubsection*{Pressure}
The pressure for a grand canonical ensemble is given by the grand canonical potential $\Omega(T,V,\mu)$.
\begin{align} \label{eq:pressure}
    P=-
    \frac{\partial \Omega (T,V,\mu)}{\partial V} \Bigg|_{T,\mu}
    =\frac{1}{\beta}\frac{g}{2\pi^2}\int_0^\kappa dp p^{2} \log(1+e^{-\beta(E(p)-\mu)})~~~.
\end{align}
where $g$ denotes the degeneracy factor, $\beta=1/T$, and $E(p)$ is the modified dispersion relation corresponding to the DSR framework under consideration. The presence of an invariant energy scale $\kappa$ restricts the allowed range of energies, leading to nontrivial modifications of the thermodynamic quantities. To study the low-temperature behavior of the system, we perform a Sommerfeld expansion of the pressure. As a first step, we integrate Eq. \eqref{eq:pressure} by parts and express the result as an integral over energy. This yields
\begin{align}
\label{eq:ibp}
    P = \frac{g}{6\pi^2}\int_{m_{0}}^{\kappa} dE \frac{\bigg[E^{2}-m^{2} \bigg(1-\frac{E}{\kappa}\bigg)^{2}\bigg]^{3/2}}{\bigg[1+\exp[\beta(E-\mu)]\bigg]}~~~.
\end{align}
The boundary terms vanish at both $p=0$ and $p\to\kappa$ since the logarithmic term vanishes, ensuring convergence of the integral despite the upper bound on the energy spectrum.
In the low-temperature regime the Sommerfeld expansion allows us to systematically extract the leading thermal corrections. The detailed derivation is presented in Appendix \ref{app:pressure_expansion}. The resulting expression for the pressure up to order $T^2$ is
\begin{align}
    P &= \frac{g}{6\pi^2} \Bigg\{\int_{m_0}^\mu \text{d} E \bigg[ E^{2} - m^{2} \bigg( 1 - \frac{E}{\kappa} \bigg)^{2} \bigg]^{3/2} + \frac{\pi^2}{4} \frac{1}{\beta^2} \Bigg[ \left[ 2 \mu + \frac{2 m^2}{\kappa} \left( 1 - \frac{\mu}{\kappa} \right)  \right] \bigg[ \mu^2 - m^2 \left( 1 - \frac{\mu}{\kappa} \right)^2 \bigg]^{1/2} \Bigg]
    +\mathcal{O}\left(\frac{1}{\beta^3}\right)\Bigg\}\,.
    \label{eq:t_0_P}
\end{align}
The first term corresponds to the zero-temperature (degenerate) pressure of the fermionic system in the DSR framework, while the second term represents the leading thermal correction proportional to $1/\beta^2$. Notably, the structure of the low-temperature expansion closely parallels that of the standard relativistic Fermi gas, indicating that the universal form of the Sommerfeld correction remains intact. The effects of DSR enter solely through the modified dispersion relation and phase-space measure. This similarity suggests that the temperature dependence of the chemical potential $\mu$ should also follow the conventional behavior. To determine the finite-temperature chemical potential, we therefore perform a Sommerfeld expansion of the number density,
\begin{align}
    n(T,\mu)
    =\int_{m_0}^\kappa dE\, \rho(E)\,n(E)\,.
\end{align}
Here $n(E)$ is the usual Fermi distribution function and $\rho(E)$ is the density of states for our theory which is the following.
\begin{align} \label{eq:distri}
    \rho(E)
    =\frac{gV}{2\pi^2}
    \Bigg[E^2-m^2\left(1-\frac{E}{\kappa}\right)^2\Bigg]^{1/2}
    \left[E
    +\frac{m^2}{\kappa}\left(1-\frac{E}{\kappa}\right)\right].
\end{align}
At finite temperature, the chemical potential no longer coincides exactly with the Fermi energy $\epsilon_f$. As in the conventional case, the leading deviation arises at order $T^2$. The low-temperature expansion of chemical potential yields
\begin{align}
    \mu
    =&\epsilon_f-\frac{\rho'(\epsilon_f)}{\rho(\epsilon_f)}\frac{1}{\beta^2}\frac{\pi^2}{6}.
\end{align}
After making all the necessary substitutions the chemical potential is obtained as
\begin{comment}
\begin{align}
    \mu=&\frac{\sqrt[3]{\pi } \left(4\ 3^{2/3} \pi ^2 m^2 n^{2/3}-\kappa^2 \left(2 \pi
   ^{2/3} m^2+4\ 3^{2/3} \pi ^2 n^{2/3}\right)\right)}{6\ 3^{2/3} \beta
   ^2 \kappa^2 n^{2/3} \sqrt{-\frac{4\ 3^{2/3} \pi ^2 m^2 n^{2/3}}{\kappa^2}+4 \pi
   ^{2/3} m^2+4\ 3^{2/3} \pi ^2 n^{2/3}}}
   +\frac{\kappa \left(\kappa \sqrt{4\
   3^{2/3} \pi ^{4/3} n^{2/3} \left(1-\frac{m^2}{\kappa^2}\right)+4 m^2}-2
   m^2\right)}{2 \left(\kappa^2-m^2\right)}\,.
\end{align}
\end{comment}
\begin{equation}
\mu=\frac{\kappa(\kappa \Lambda -m^2)}{\kappa^2 - m^2} - \frac{\pi^{2} \left[m^2 + 2p_F^2 \left(1-\frac{m^2}{\kappa^2}\right)\right]}{6 \beta^2 p_F^2 \Lambda}\,,
\end{equation}
with the relatistic fermi momentum as 
\begin{align}
    p_F^2=\frac{1}{4} \frac{6^{2/3}}{\pi^{2/3}} \left(\frac{h^3 n}{g}\right)^{2/3}\,,\label{eq:fermi_mom}
\end{align}
and a quantity $\Lambda$ as 
\begin{align}
    \Lambda = \sqrt{m^2 + p_F^2 \left(1-\frac{m^2}{\kappa^2}\right)}\,.\label{eq:Lambda}
\end{align}
We use the condition that the fluctuations of number density at low-temperature is negligible. It appears in the chemical potential through the Fermi energy, which for our DSR is as follows.
\begin{align}
    \epsilon_f
    =\frac{1}{1-\frac{m^2}{\kappa^2}}
    \Bigg[
    -\frac{m^2}{\kappa}
    +\sqrt{m^2+\bigg(1-\frac{m^2}{\kappa^2}\bigg)\bigg(\frac{6\pi^2}{g}n\bigg)^{2/3}}\Bigg]
\end{align}
This expression explicitly reflects the interplay between relativistic kinematics and the invariant energy scale $\kappa$, and it reduces smoothly to the standard relativistic result in the limit $\kappa\to\infty$.
\subsubsection*{Entropy}
The entropy of the system in the grand canonical ensemble can be obtained directly from the grand potential $\Omega(T,V,\mu)$ through the standard thermodynamic relation
\begin{align} \label{eq:entropy}
    S = -\frac{\partial \Omega(T,V,\mu)}{\partial 
    T}\Bigg|_{V,\mu}\,.
\end{align}
This definition allows us to evaluate the entropy once the temperature dependence of the grand potential is known. In the present framework, the modified dispersion relation associated with doubly special relativity (DSR) introduces nontrivial corrections to $\Omega$, and consequently to the entropy, through the deformed phase-space structure and the presence of an invariant energy scale $\kappa$. We are primarily interested in the behavior of the entropy in the low-temperature regime, where $T$ is small but finite. In this limit, the Sommerfeld expansion provides a controlled approximation, allowing us to express thermodynamic quantities as a series in powers of temperature. Carrying out this expansion for the entropy, we obtain
\begin{align}
\label{eq:S_T->0}
    S =& \frac{gV}{2\pi^2} \bigg\{\int_{m0}^\mu \text{d}E
    \bigg[\frac{\beta}{3} \left(E^2-m^2\bigg(1-\frac{E}{\kappa}\bigg)^2\right)^{3/2}
    +\beta\left(E^2-m^2\bigg(1-\frac{E}{\kappa}\bigg)^2\right)^{1/2}
    \left(E+\frac{m^2}{\kappa}\bigg(1-\frac{E}{\kappa}\bigg)\right)
    (E-\mu)\bigg]\nonumber\\
    &+\frac{\pi^2}{3}
    \frac{1}{\kappa^2\beta}(\kappa m^2+\kappa^2\mu-m^2\mu)
    \bigg(
    \mu^2-\frac{m^2}{\kappa^2}(\kappa-\mu)^2
    \bigg)^{1/2}
    +\mathcal{O}\left(\frac{1}{\beta^3}\right)
    \bigg\}.
\end{align}
The first term, corresponding to the entropy at $T=0$, actually is evaluated to 0. This ensures that the entropy at absolute zero temperature is 0. This result is consistent with the third law of thermodynamics and confirms that the DSR framework considered here does not introduce any residual entropy at absolute zero. It is important to note that, unlike the pressure and chemical potential, whose leading thermal corrections appear at order $T^2$, the entropy of a degenerate Fermi system is linear in temperature at low $T$. This behavior is clearly reflected in Eq. \eqref{eq:S_T->0}, where the leading nonvanishing contribution to the entropy is proportional to $1/\beta$. Physically, this linear dependence arises from thermal excitations of particles in a narrow energy shell around the Fermi surface, whose width is of order $T$. As mentioned earlier, our analysis focuses on thermodynamic quantities up to order $1/\beta^2\sim T^2$. One might therefore expect that higher-order terms in the Sommerfeld expansion could contribute additional corrections to the entropy at this order. However, a careful examination shows that the next term in the expansion, which would nominally appear at order $1/\beta^2$, vanishes identically for the present system. Consequently, no additional contributions arise at this order, and Eq. \eqref{eq:S_T->0} already represents the complete expression for the entropy within our approximation scheme.
\subsubsection*{Internal energy}
In the grand canonical ensemble, the internal energy $U$ of a system is related to the partition function through the thermodynamic identity
\begin{align}
    U = \int_0^\kappa\, \text{d}p\,p^2\,E\,\frac{1}{1+e^{\beta(E-\mu)}}.
\end{align}
The integrand in this expression corresponds to the energy-weighted distribution of particles in momentum space, with the Fermi-Dirac distribution $\frac{1}{1+e^{\beta(E-\mu)}}$ determining the occupation of states at each energy level. By following the same steps used in the analysis of the pressure and entropy, we arrive at the following expression for the internal energy to order $T^2$.
\begin{align}
    U=&\frac{gV}{2\pi^2}
    \bigg\{\int_{m0}^\mu dE
    \left(E^2-m^2\bigg(1-\frac{E}{\kappa}\bigg)^2\right)^{1/2}
    \left(E+\frac{m^2}{\kappa}\bigg(1-\frac{E}{\kappa}\bigg)\right)
    E\nonumber\\
    &+\frac{1}{\beta^2}
    \frac{\pi^2}{6}
    \frac{-7 \kappa m^4 \mu^2 + 3 m^4 \mu^3 - \kappa^3 (m^4 - 7 m^2 \mu^2) + 
 \kappa^4 (-2 m^2 \mu + 3 \mu^3) + 
 \kappa^2 (5 m^4 \mu - 6 m^2 \mu^3)}
{\kappa^4 \sqrt{-\left(\frac{m^2 (\kappa - \mu)^2}{\kappa^2}\right) + \mu^2}}
+\mathcal{O}\left(\frac{1}{\beta^3}\right)
    \bigg\}~~.
    \label{eq:t_0_E}
\end{align}
The first term in this expansion corresponds to the internal energy at zero temperature, which, as in the cases of pressure and entropy, arises from the degenerate fermionic gas. The modification introduced by the DSR framework is apparent in the altered dispersion relation, which includes a maximum energy scale $\kappa$ that truncates the available states. This ensures that the internal energy is bounded, even at extremely high densities or energies. At finite temperatures, the second term in Eq. \eqref{eq:t_0_E} represents the leading thermal correction, which scales as $T^2$(or equivalently, $1/\beta^2$). The higher-order corrections, which would appear at order $T^4$, are not included in this analysis.

\begin{figure}
    \centering
    \includegraphics[width=0.49\linewidth]{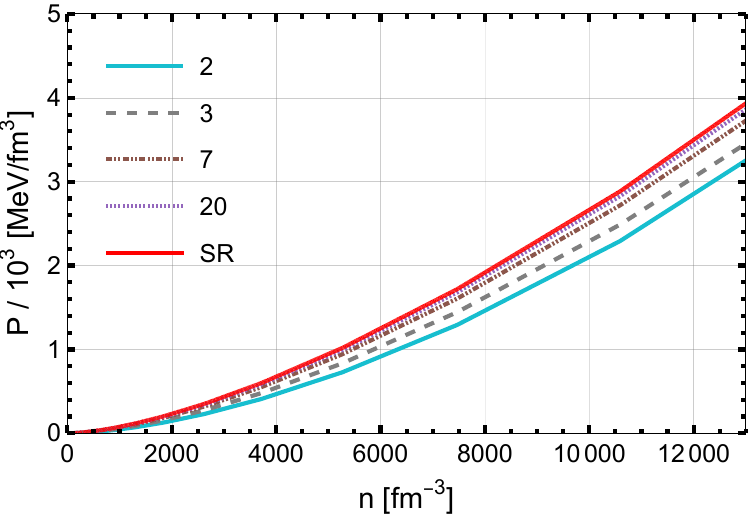}
    \includegraphics[width=0.49\linewidth]{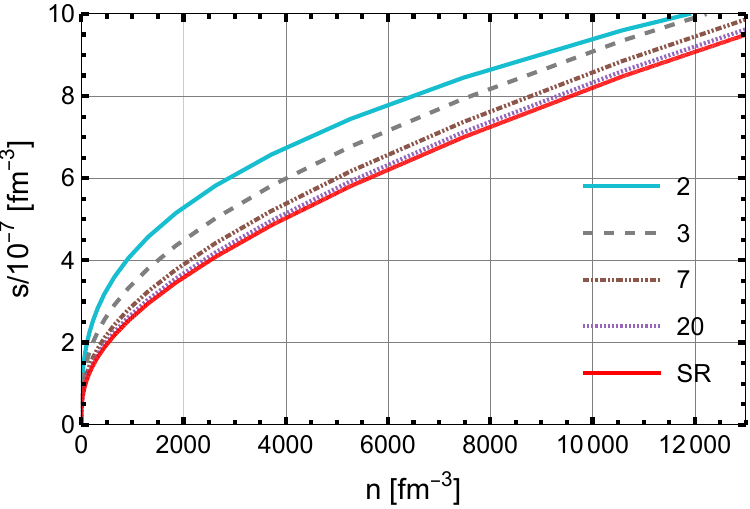}
    \includegraphics[width=0.49\linewidth]{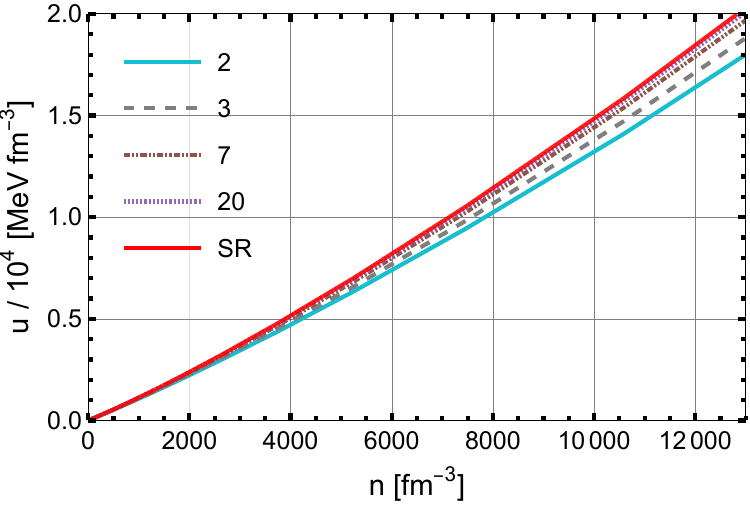}
    \includegraphics[width=0.49\linewidth]{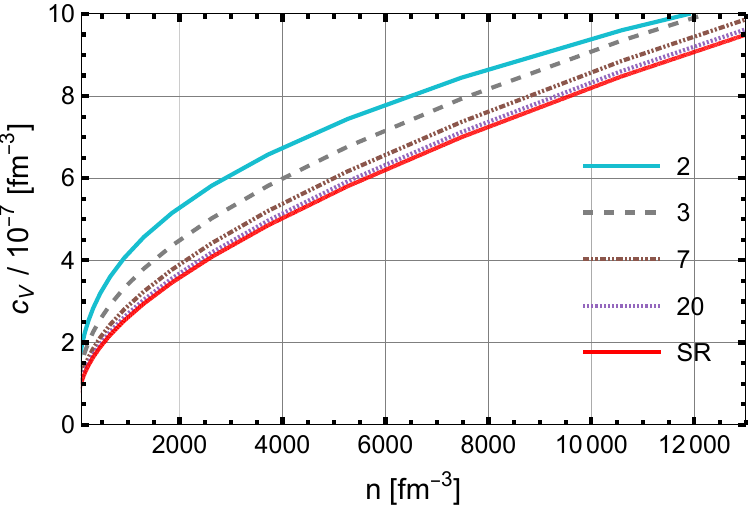}
    \caption{Pressure $(P)$, entropy $(s=S/V)$, internal energy $(u=U/V)$ and specific heat $(c_V=C_V/V)$ vs. particle number density (n)  for $\kappa = 2\, \mathrm{\mathrm{GeV}},\, 3\, \mathrm{\mathrm{GeV}},\, 7\, \mathrm{\mathrm{GeV}},\, 20\, \mathrm{\mathrm{GeV}}$ DSRs in the low-temperature limit. The solid red curves represent the thermodynamic quantities in the special relativistic case, i.e., in the limit $\kappa\to\infty$. These results are from taking a low-temperature of $8.62\times10^{-9} \,\mathrm{\mathrm{GeV}}$ and a rest mass of $0.939$ GeV.}
    \label{fig:low-temp}
\end{figure}

\subsubsection*{Specific heat}
The specific heat at constant volume, $C_V$, describes how the internal energy $U$ of a system changes with respect to temperature. In the grand canonical ensemble, the specific heat is given by the thermodynamic relation
\begin{align} \label{eq:specificheat}
    C_V=\frac{\partial U}{\partial T}\Bigg|_{V,N}\,.
\end{align}
In the low-temperature limit, the internal energy given by Eq. \eqref{eq:t_0_E} is expanded in powers of $T$, with the leading term corresponding to the zero-temperature energy of the system and the next term capturing the finite-temperature correction at order $T^2$. Taking the derivative of the internal energy with respect to temperature, we obtain the specific heat at low- temperature.
\begin{align}
    C_V=&\frac{gV}{2\pi^2}\Bigg\{\frac{k_B}{\beta}\frac{\pi^2}{3}\,\frac{-7\kappa m^4\mu^2+3m^4\mu^3-\kappa^3(m^4-7m^2\mu^2)+\kappa^4(-2m^2\mu+3\mu^3)+\kappa^2(5m^4\mu-6m^2\mu^3)}{\kappa^4\sqrt{-\left(\frac{m^2(\kappa-\mu)^2}{\kappa^2}\right)+\mu^2}}\nonumber\\
&-\,k_B\beta^2\,\frac{\partial\mu}{\partial\beta}\left(\mu^2-m^2\left(1-\frac\mu\kappa\right)^2\right)^{1/2}\left(\mu+\frac{m^2}\kappa\left(1-\frac\mu\kappa\right)\right)\mu \;+\;\mathcal{O}\!\left(\frac1{\beta^3}\right)\Bigg\}
\end{align}
The first term in this expression represents the leading contribution to the specific heat at low-temperature, capturing the effect of thermal fluctuations on the system. This term is proportional to $1/\beta$, reflecting the typical temperature dependence for systems where the excitation spectrum is dominated by fermions near the Fermi surface. 

The plots in Fig. \ref{fig:low-temp} show the variation of thermodynamic quantities as a functions of the number density for fermions with mass equal to that of the neutron. In this regime, the temperature is much smaller than the particle rest mass (expressed in $\mathrm{GeV}$ units), justifying the use of the low-temperature limit. From the plots, it is evident that for $\kappa=2\,\mathrm{GeV}$, the thermodynamic quantities deviate significantly from those obtained using the standard special relativistic (SR) dispersion relation. As $\kappa$ increases, the DSR results approach the SR curve, indicating recovery of the standard relativistic behavior in the large-$\kappa$ limit. As can be seen, for $\kappa=20\,\mathrm{GeV}$, the DSR predictions are very close to the SR case. Therefore, for the chosen physical parameters, values of $\kappa\lesssim10\,\mathrm{GeV}$ lead to appreciable deviations from SR and represent the regime where DSR effects become significant.
\subsection{At high-temperature} \label{sec:thermoT=inf}
To investigate the thermodynamics of the Fermi gas in the high-temperature regime, we begin with the grand-canonical partition function. Rather than the partition function itself, it is convenient to work with its logarithm,
\begin{equation}
    \log \mathcal{Z} = \frac{V}{2\pi^2}g \int_{m_{0}}^{\kappa} dE\, \left[E^{2}-m^{2}\bigg(1-\frac{E}{\kappa}\bigg)^{2}\right]^{1/2}\left[E+\frac{m^{2}}{\kappa}\bigg(1-\frac{E}{\kappa}\bigg)\right]\log\left[1+e^{-\beta(E-\mu)}\right]. 
\end{equation}
We can rewrite this as follows:
\begin{align}
    \log \mathcal{Z} & =  \frac{V}{2\pi^2}g \int_{m_{0}}^{\kappa} dE\, \rho(E)\,\log\left[1+e^{-\beta(E-\mu)}\right],
\end{align}
where $\rho(E) = \left[E^{2}-m^{2}\bigg(1-\frac{E}{\kappa}\bigg)^{2}\right]^{1/2}\left[E+\frac{m^{2}}{\kappa}\left(1-\frac{E}{\kappa}\right)\right]$. 
In the high-temperature regime, the standard low- temperature techniques based on expanding the Fermi Dirac distribution are no longer applicable. In particular, when the temperature is large, the exponential factor in the distribution function cannot be treated perturbatively. A controlled approximation therefore requires a careful identification of the relevant hierarchy among the physical scales appearing in the theory. The thermodynamics of the model is governed by three parameters: the particle mass $m$, the temperature $T$, and the ultraviolet energy scale $\kappa$ associated with the modified dispersion relation. The high-temperature regime of interest corresponds to energies that are large compared to the rest mass but remain well below the ultraviolet cutoff. This can be presented as follows:
\begin{align} \label{eq:relation}
m \ll E \sim T \ll \kappa
\end{align}

\begin{figure}
    \centering
    \includegraphics[width=0.49\linewidth]{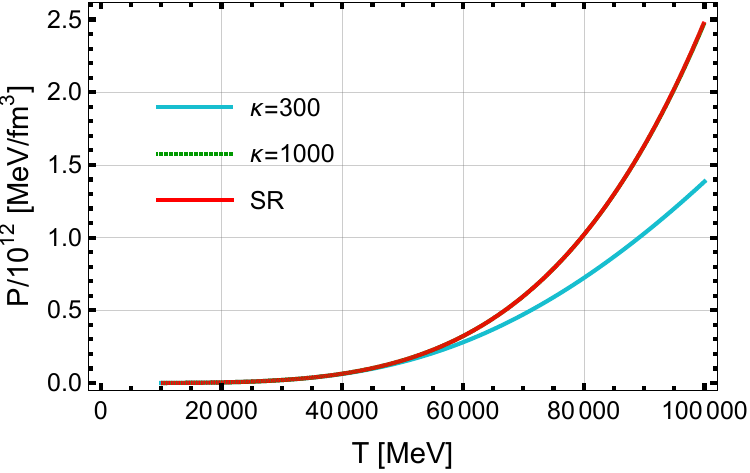}
    \includegraphics[width=0.49\linewidth]{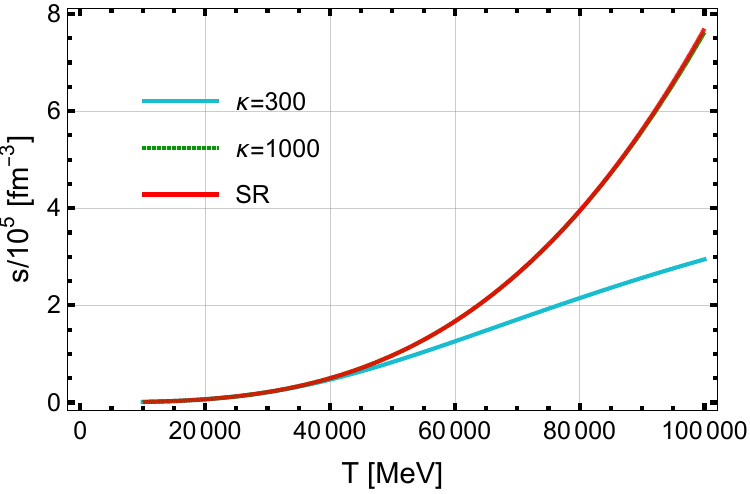}
    \includegraphics[width=0.49\linewidth]{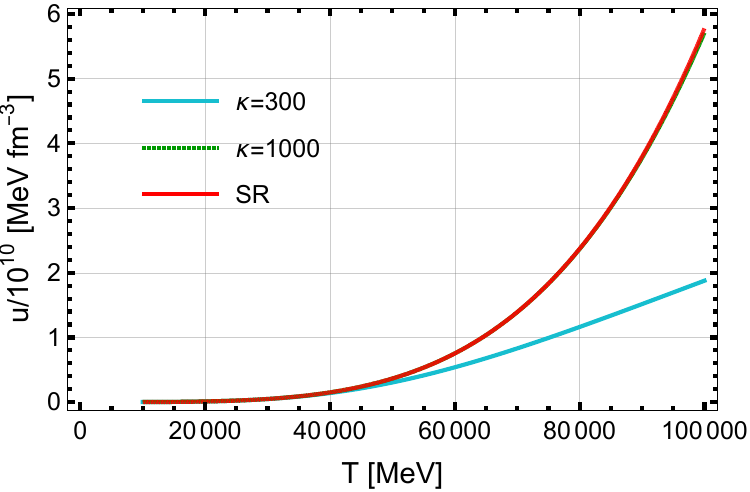}
    \includegraphics[width=0.49\linewidth]{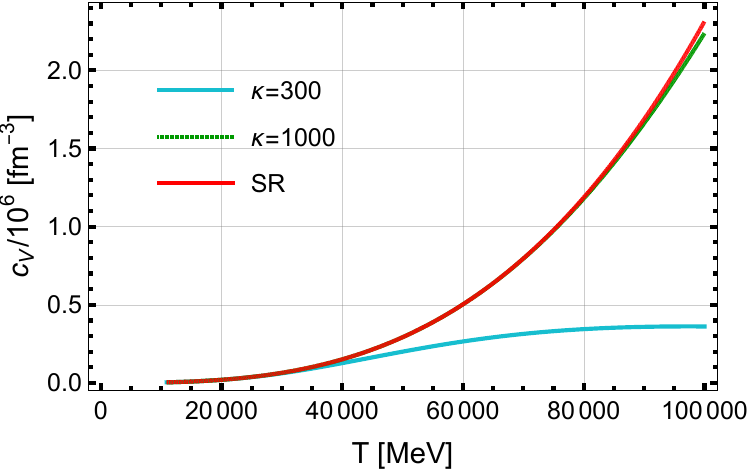}
    \caption{The pressure $(P)$, entropy $(s=S/V)$, internal energy $(u=U/V)$, and specific heat $(c_V=C_V/V)$ are plotted as functions of the temperature $(T)$ at fixed number density in the high-temperature regime. The blue curves correspond to the deformed case with $\kappa = 300~\mathrm{\mathrm{GeV}}$ and $\kappa = 1000~\mathrm{GeV}$, while the red curves represent the standard relativistic limit, $\kappa \rightarrow \infty$. The plots are obtained by taking the rest mass to be $m_{0} = 0.939~\mathrm{\mathrm{GeV}}$ and imposing the constraint $m \ll T \ll \kappa$.}
    \label{fig:high-temp}
\end{figure}

In this limit, relativistic effects dominate over mass corrections, while the ultraviolet deformation remains perturbative. As a consequence, the dimensionless ratios $\frac{m^{2}}{E^{2}}$ and $\frac{E}{\kappa}$ are both small, providing a natural expansion scheme. In the high-temperature regime, two physically distinct situations arise. In the first case, the gas has no net conserved charge — pair creation is unsuppressed and $\mu = 0$ follows from charge neutrality, as derived below. This reflects the fact that thermal excitations dominate over quantum degeneracy effects. In second case, the particle number density $n$ is fixed, as in compact star matter, and $\mu(T,n,\kappa)$ is obtained by solving the number equation numerically; the qualitative behaviour of $\mu$ in this regime is discussed in Appendix \ref{sec:appB}. The present section focuses exclusively on the first case. Figure \ref{fig:high-temp} presents the thermodynamic quantities in the first case.

For a gas with no conserved charge, the particle number is not fixed because particle-antiparticle pairs can be created and annihilated. In this case, the chemical potential is not determined by minimizing the grand potential with respect to $N$, since
\begin{align}
    \left(\frac{d\Omega}{d\mu}\right)_{T,V} = 0,
\end{align}
and setting this derivative to zero would imply $\langle N \rangle = 0$, which is not the physical condition of interest. Instead, the correct argument follows from particle-antiparticle symmetry. For a relativistic gas in thermal equilibrium, the net conserved charge is
\begin{align}
    Q & = \langle N_{+} \rangle - \langle N_{-} \rangle = -\frac{g\,V}{2\pi^{2}}\, \int_{m_{0}}^{\kappa}\,dE\, \rho(E)\,\bigg[\log{\left[1+e^{-\beta\,(E-\mu)}\right]-\log{\left[1+e^{-\beta\,(E+\mu)}\right]}}\bigg],
\end{align}
In the absence of a conserved charge, the equilibrium condition is $Q = 0$. Since the integrand is an odd function of $\mu$, this condition is satisfied only for
\begin{align}
    \mu & = 0
\end{align}
We therefore consider a gas with no conserved charge and unrestricted particle-antiparticle pair creation, for which the chemical potential vanishes, $\mu= 0$. All thermodynamic quantities in the first case are evaluated under this condition. The partition function is given by
\begin{align}
    \log \mathcal{Z} & =\frac{V\, g}{2\pi^2}\int_{m_{0}}^{\kappa} dE\, \rho(E)\, \log\left[1+e^{-\beta E}\right]. 
\end{align}
Using $m = \frac{m_{0}}{\left(1-\frac{m_{0}}{\kappa}\right)}$ and the grand potential $(\Omega)$ is obtained using
\begin{align}
    \Omega(T, V, \mu = 0) & = - T\,\log[\mathcal{Z}(T,V,\mu = 0)], 
\end{align}
In the high-temperature, the thermodynamic quantities are obtained from energy integrals involving the modified density of states. In order to make analytic progress, it is tempting to expand this density in powers of $\frac{m}{E}$ and $\frac{E}{\kappa}$ and truncate the resulting series at leading order. However, such a truncation is only asymptotic and is not uniformly valid over the full integration domain. In particular, while the expansion is controlled for energies satisfying $m \ll E \ll \kappa$, it breaks down at low energies and near the ultraviolet cutoff, where the exact density of states remains positive but its truncated form may become negative. As a consequence, the truncated expressions can lead to unphysical artifacts in the thermodynamic functions, such as negative pressure or entropy, signaling a breakdown of the approximation rather than a genuine physical effect. This limitation of the truncated expansion must therefore be kept in mind when interpreting analytic high-temperature results. To avoid these inconsistencies and to ensure thermodynamic stability, we therefore evaluate all thermodynamic quantities directly from the exact expressions by performing the energy integrals numerically with the constraint $m \ll E \ll \kappa$. In the following, we present numerical results for a representative finite value of the deformation scale $\kappa$ and compare them with the case $\kappa \to \infty$, which corresponds to the standard relativistic (SR) dispersion relation. In the high-temperature regime $m \ll T \ll \kappa$, the two cases are found to be very close, as expected, since the effects of the deformation are suppressed by powers of $\frac{T}{\kappa}$. The pressure $(P)$, entropy $(S)$, internal energy $(U)$, and specific heat $(C_V)$ in the grand canonical ensemble with vanishing chemical potential $(\mu = 0)$ are obtained using the following relations:
\begin{align} \label{eq:thermorelationshigh}
P &= -\frac{\Omega}{V}, \\
S &= -\left(\frac{\partial \Omega}{\partial T}\right)_{V}, \\
U &= \Omega + T\cdot S, \\
C_V &= \left(\frac{\partial U}{\partial T}\right)_{V}.
\end{align}
All thermodynamic quantities in the high-temperature regime are evaluated numerically from the exact expressions for the grand canonical potential, without invoking any truncated expansion of the density of states. This procedure avoids the unphysical artifacts that can arise from non uniform high-temperature expansions and ensures thermodynamic consistency. The numerical analysis is performed for a physical mass $m_0 = 0.939~\mathrm{\mathrm{GeV}}$ and for temperature in the range $20 \leq T \leq 100~\mathrm{\mathrm{GeV}}$. We consider a representative finite value of the deformation scale $\kappa$, and compare the results with the undeformed limit $\kappa \to \infty$, which corresponds to the standard relativistic (SR) case. As $\kappa$ increases, the effective mass $m = \frac{m_{0}}{1-\frac{m_{0}}{\kappa}}$ approaches $m_0$, and the deformed theory smoothly converges to the SR limit.

The numerical results in Fig. \ref{fig:high-temp} show that the overall temperature dependence of all thermodynamic observables is similar in the deformed and undeformed cases within the range considered, reflecting the fact that $T \ll \kappa$ and that deformation effects are therefore parametrically suppressed. Nevertheless, systematic numerical differences are observed. In the deformed case the pressure and internal energy are reduced relative to the SR case at the same temperature, indicating a softening of the equation of state and a suppression of the contribution of high energy modes. The entropy likewise grows more slowly with temperature, indicating a reduced rate of increase of the number of accessible microstates due to the presence of the invariant scale. Correspondingly, the specific heat is suppressed and tends to flatten at higher temperature, indicating that the system becomes progressively less efficient at absorbing thermal energy as the contribution of high energy states is limited. These effects provide a consistent thermodynamics of the role played by the invariant energy scale $(\kappa)$ in regulating the high energy behavior of the system, while preserving smooth convergence to the standard relativistic limit as $\kappa \to \infty$. 

\section{Astrophysical application}
\label{sec:astrophysics}
In this section we apply the modified thermodynamic framework to compact stellar objects whose equilibrium is governed by Fermionic degeneracy pressure. In particular, we focus on stellar objects in which the degeneracy pressure provides the dominant support against gravitational collapse. The presence of an invariant energy scale $(\kappa)$ alters the equation of state of the degenerate matter and can therefore modify the standard mass–radius relations. We analyze these effects within a simplified but physically transparent model, considering both white dwarf stars and neutron stars as representative examples, and briefly discuss the phenomenological implications of the resulting modifications.
\subsection{White dwarf}
\label{subsec:white-dwarf}
A standard model of a white dwarf star consists of $N$ free electrons and $\frac{N}{2}$ free helium nuclei \cite{chandrasekhar1957introduction}. The mass of the particles in the star is given by $(m_{n}\approx m_{p})$
\begin{align} \label{eq:stellarMass}
    M = Nm_{e} + \frac{N}{2}(2m_{n}+2m_{p})\approx Nm_{e} + 2Nm_{p} = N(m_{e}+2m_{p})
\end{align}
here $m_{e}$ and $m_{p}$ represent the rest masses of the electron and proton, respectively. The pressure supporting the star against gravity is provided by a gas of degenerate electrons, while the mass density is mainly due to non-degenerate ions such as of carbon or helium \cite{padmanabhan2006invitation,padmanabhan2000theoretical,sobolev1969course, 2010A&ARv..18..471A}. The contribution of the nuclei to the pressure is negligible and can be ignored, since they mainly affect the mass of the star and not its mechanical support. The typical internal temperature of a white dwarf is of order $10^{7}K$, which is too small to prevent gravitational collapse by thermal pressure alone. On the other hand, the Fermi energy of the electrons is of order $10^{9} K$, much larger than the thermal energy scale. This ensures that the electrons are highly degenerate, and the electron gas can be treated as a zero temperature Fermi gas to a very good approximation. We consider the limit $T\rightarrow 0$ rather than strictly $T = 0$ in order to include the leading thermal corrections to the degenerate equation of state. In this limit, the contribution of radiation can also be neglected, since almost all electron states up to the Fermi energy are occupied. In what follows, we focus on the ultra relativistic regime in which the electron momentum satisfies $cp\gg mc^{2}$. The degenerate pressure $(P_{g})$ is obtained in the previous section (\ref{sec:T=0}) and can be written in the polytropic form \cite{chandrasekhar1957introduction}
\begin{align} \label{eq:polytrope}
    P_{g} & = K \rho^{\gamma},
\end{align}
where $K$ is a constant determined by the microscopic equation of state, $\rho$ is the mass density, and $\gamma = 1+\frac{1}{n}$ defines the polytropic index $n$. In the present model both $K$ and $n$ depend on the invariant energy scale $\kappa$ through the modified dispersion relation (Eq. \eqref{eq:modified}), and are obtained by conducting a linear least-squares regression on log P vs. log $\rho$.

We model the white dwarf as a static, spherically symmetric configuration in hydrostatic equilibrium. The balance between the outward degeneracy pressure and the inward gravitational force is governed by
\begin{align}
\frac{dP}{dr} = -\frac{G M(r)}{r^2}\rho(r),
\end{align}
where $M(r) = 4\pi\, \int_{0}^{r}\, dr'\, r'^{2}\,\rho(r')$ is the mass enclosed within radius $r$. Combining this equation with the polytropic equation of state yields
\begin{align}\label{eq:poly1}
    \frac{1}{r^{2}}\frac{d}{dr}\left(\frac{r^{2}}{\rho}\frac{dP_{g}}{dr}\right) = -4\pi G \rho(r)
\end{align}
Introducing the dimensionless variables
\begin{align}
\rho(r) & = \rho_0 \Theta^n(\xi),\hspace{2em}
r  = \alpha \xi,\qquad
\alpha = \left[\frac{(n+1)K\rho_0^{(1-n)/n}}{4\pi G}\right]^{1/2},
\end{align}
and then Eq. \eqref{eq:poly1} reduces to the Lane Emden equation \cite{padmanabhan2006invitation,shapiro2024black,chandrasekhar1957introduction},
\begin{align}
    \frac{1}{\xi^{2}} \frac{d}{d\xi}\left(\xi^{2}\frac{d\Theta}{d\xi}\right) = - \Theta^{n}~~~.
    \label{lane-emden}
\end{align}
The boundary conditions are $\Theta(0) = 1$ and $\Theta'(0) = 0$, corresponding to a finite central density and vanishing density gradient at the center. The stellar surface is defined by the first zero $\xi = \xi_{1}$ of $\Theta$ and the radius of the star $(R)$ is therefore
\begin{align}
 R & = \alpha \xi_{1} = \left(\frac{(n+1)K\rho_{0}^{(1-n)/n}}{4\pi G}\right)^{1/2} \xi_{1}.
 \label{wd-radius}
\end{align}
The total mass $(M)$ is obtained by integrating the density profile,
\begin{align}
    M & = 4\pi\int_{0}^{R}dr\, r^{2}\rho(r) \label{eq:total-mass} \\
    &= 4\pi\left(\frac{(n+1)K}{4\pi G}\right)^{3/2} \rho_{0}^{(3-n)/2n} \xi^{2}_{1}~|\Theta'(\xi_{1})|. 
\end{align}
Eliminating the central density $\rho_{0}$ between the expressions for $M$ and $R$ yields the mass–radius relation
\begin{align}
    M = 4\pi\left(\frac{(n+1)K}{4\pi G}\right)^{n/(n+1)}\xi_{1}^{(n+3)/(1-n)}\xi_{1}^{2}~|\Theta'(\xi_{1})|~R^{(n+3)/(1-n)}.
    \label{wd-mass}
\end{align}

\begin{figure}[tbp]
    \centering
    \includegraphics[width=0.48\linewidth]{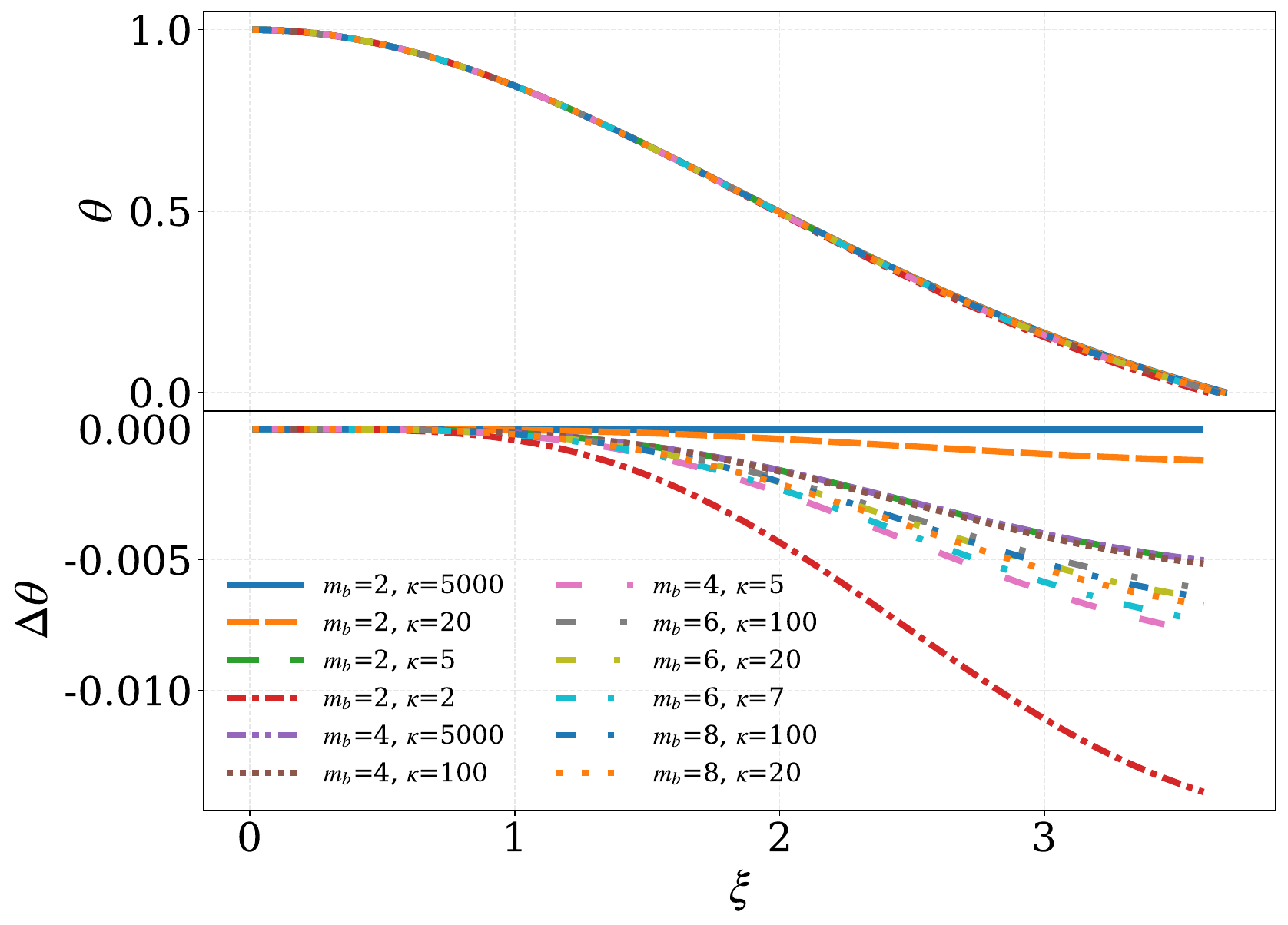}
    \includegraphics[width=0.48\linewidth]{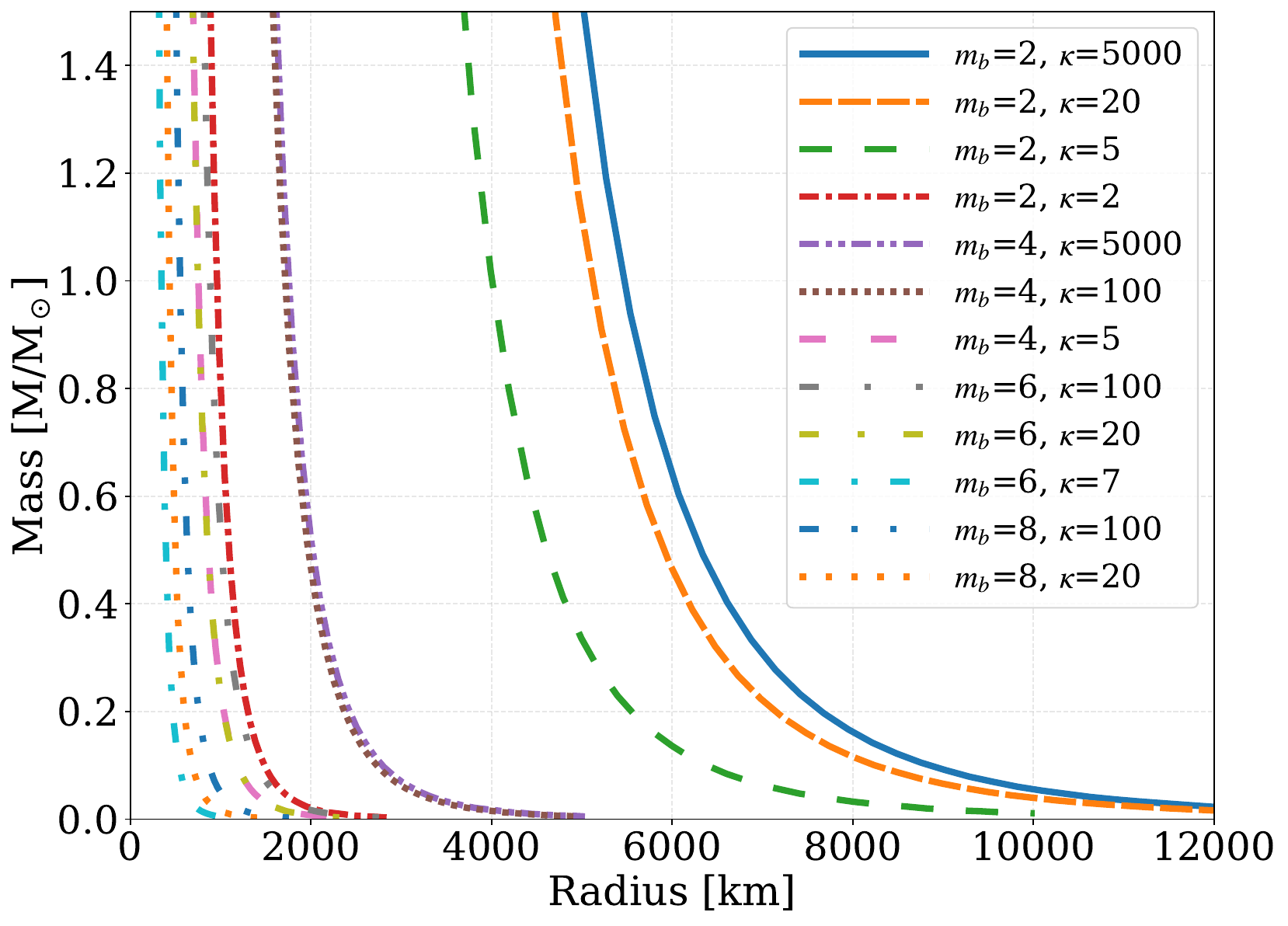}
    \caption{\textit{Left}: Lane-Emden parameters $\theta$ vs. $\xi$ for combinations of $m_b$ and $\kappa$. Lower panel shows $\Delta\theta$ (w.r.t. $m_b=2$ and $\kappa=5000~\rm{\mathrm{GeV}}$ curve) vs. $\xi$ to highlight differences in the $\theta-\xi$ curves. \textit{Right}: White dwarf mass-radius produced by the DSR for a range of particle mass $m_b$ and $\kappa$.}
    \label{fig:wd-MR}
\end{figure}
Stellar burning produces three classes of white dwarfs: He, CO, or ONeMg. The elements in the class-names refer to the elements present in the core of the white dwarf, produced due to increasing levels of burning undergone in the progenitor star. Lighter stars ($M\lesssim$ 0.5 M$_\odot$) are unable to ignite He and lead to He white dwarf. More heavier stars undergo more elemental burning to synthesize up to C and O (0.5 M$_\odot\lesssim M\lesssim$ 8 M$_\odot$) or O, Ne, and Mg (8 M$_\odot\lesssim M\lesssim$ 11 M$_\odot$). Stars heavier than this lead to further fusion up to Fe and cause either a successful or a failed core-collapse supernova that leave a remnant neutron star and a black hole respectively. Later in Section \ref{subsec:neutron-star} we discuss the implications on neutron stars structure due to DSR.

To model the various possible compositions of white dwarf cores, we assume mean baryon mass $m_{b} = 2~m_n,~4~m_n,\allowbreak ~ 6~m_n, \rm{and}~8~m_n$, solve the Lane-Emden equation, Eq. \eqref{lane-emden}, for $\xi_1$ and use Eqs. \eqref{wd-radius} and \eqref{wd-mass} within the DSR constraints. Most of the mass of white dwarfs is composed of symmetric nuclei ($Z=N$) such as $^4$He, $^{12}$C, $^{16}$O, $^{20}$Ne, or $^{24}$Mg. Thus, the mean molecular weight per electron for all the white dwarfs considered is $\approx$ 2 amu \cite{2010A&ARv..18..471A}. Left panel of Fig. \ref{fig:wd-MR} shows the solutions for the Lane-Emden equation Eq. \eqref{lane-emden} obtained for polytropic index $n$ for combinations of $m_b$ and $\kappa$. For each value of $m_b$ the $\Delta\theta$ is larger for larger $\kappa$ while for a constant $\kappa$, $\Delta\theta$ is larger for smaller $m_b$. This results in larger radii for larger $\kappa$ and smaller $m_b$ as can be seen in the right panel Fig. \ref{fig:wd-MR} which shows the mass-radius relations for the same $m_{b}$ and $\kappa$ combinations. The radii obtained for low $m_b$ and high $\kappa$ combinations are compatible with general Lane-Emden solutions, with lowest $m_b = 2~m_n$ and highest $\kappa=5000\,\mathrm{GeV}$ producing the largest radii. Larger $\kappa$ approximates the DSR to standard SR because $E/\kappa \rightarrow 0$ as discussed in Sec. \ref{Sec:Moddispersion}; hence this suppresses the contribution of the modification in the DSR. However, we also note a unique behavior for higher $m_b$ where the differences between M-R for the range of $\kappa$ values are minimal. For instance for $m_b=8~m_n$, $\kappa=20\, \mathrm{GeV}$ and $\kappa=100\, \mathrm{GeV}$ lead to similar mass-radius. Similarly,  $m_b=2~m_n$, $\kappa=2~\mathrm{\mathrm{GeV}}$ and $m_b=6~m_n$, $\kappa=100~\mathrm{\mathrm{GeV}}$, and $m_b=4~m_n$, $\kappa=5000~\mathrm{\mathrm{GeV}}$ and $m_b=4~m_n$, $\kappa=100~\mathrm{\mathrm{GeV}}$ are indistinguishable. This degeneracy between various DSR parameter $\kappa$ and white dwarf component $m_b$, indicates the possibility of DSR fitting white dwarf observational data with modified or mixed composition.

\subsection{Neutron star}
\label{subsec:neutron-star}

Neutron stars are compact remnants of successful core-collapse supernovae of massive stars ($M\gtrsim 10 M_\odot$). After cooling down from the collapse, neutron stars remain in a hydrostatic equilibrium where the gravitational pressure pulling the nucleons is balanced by the repulsive strong force and degeneracy pressure. Observations of spinning neutron stars have been made for over decades, however, recently several unique and insightful observations of neutron stars have been made. The observation of high mass neutron star observations ($M\gtrsim 2M_\odot$) \cite{Antoniadis:2013pzd}, radius constraints on neutron stars via NICER X-ray pulse profiling \cite{Miller:2019cac, Riley:2019yda, Miller:2021qha, Riley:2021pdl}, neutron star mergers via gravitational waves by the LIGO-Virgo collaboration \cite{LIGO-GW170817-EOS}, as well as a low mass-low radius measurement HESS J1731-347 \cite{HESS}, cover distinct techniques and neutron star populations. Incorporating all observations under one equation of state of nucleonic interaction has been found challenging \cite{kedia2024} suggesting a strong \textit{soft} to very \textit{stiff} shift at a few times nuclear saturation density, which nucleonic models may not be able to attain. This suggests potential modifications to underlying theory as a way to explain current challenges.

\begin{figure}[tbp]
    \centering
    \includegraphics[width=0.44\linewidth]{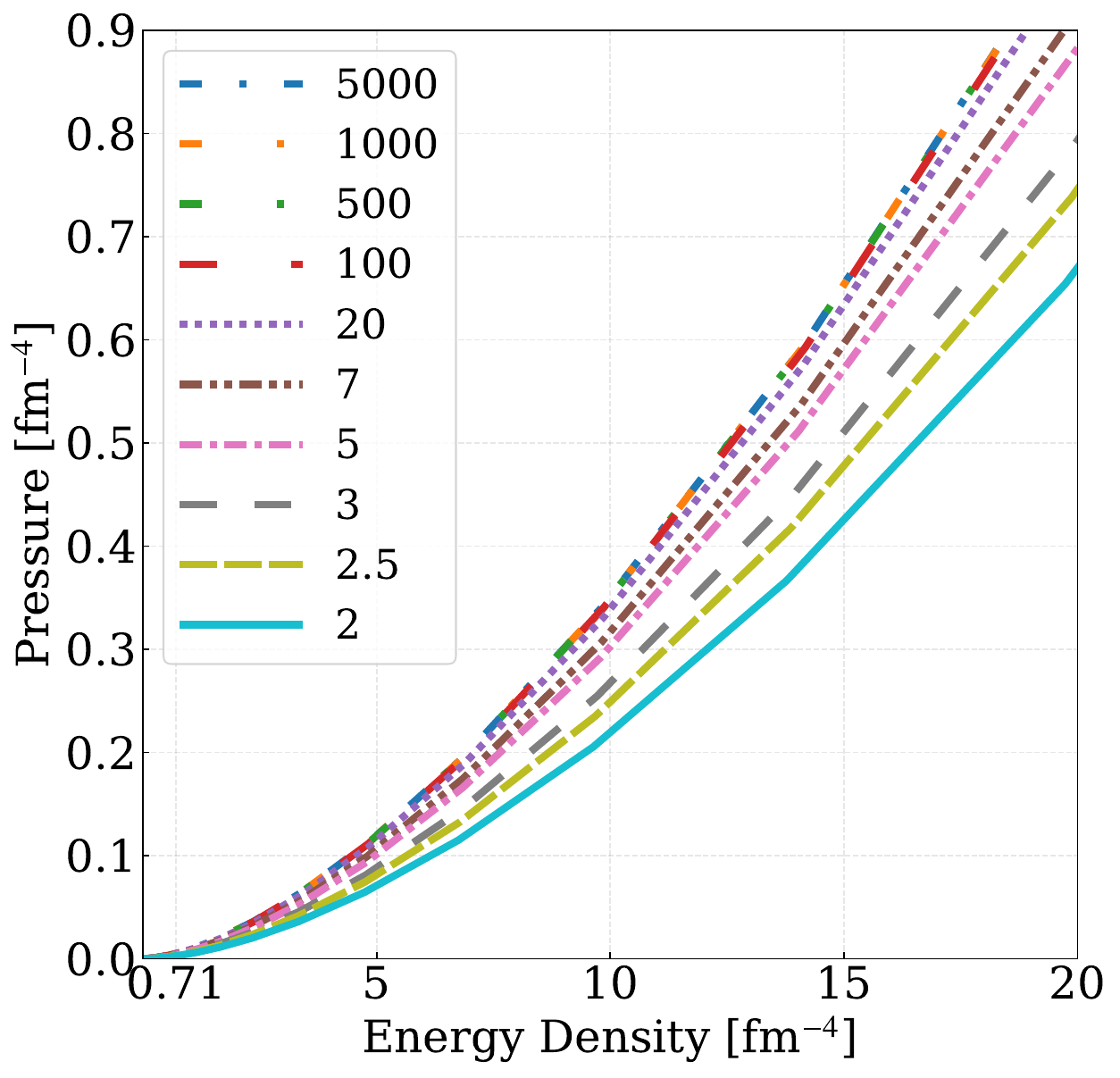}
    \includegraphics[width=0.55\linewidth]{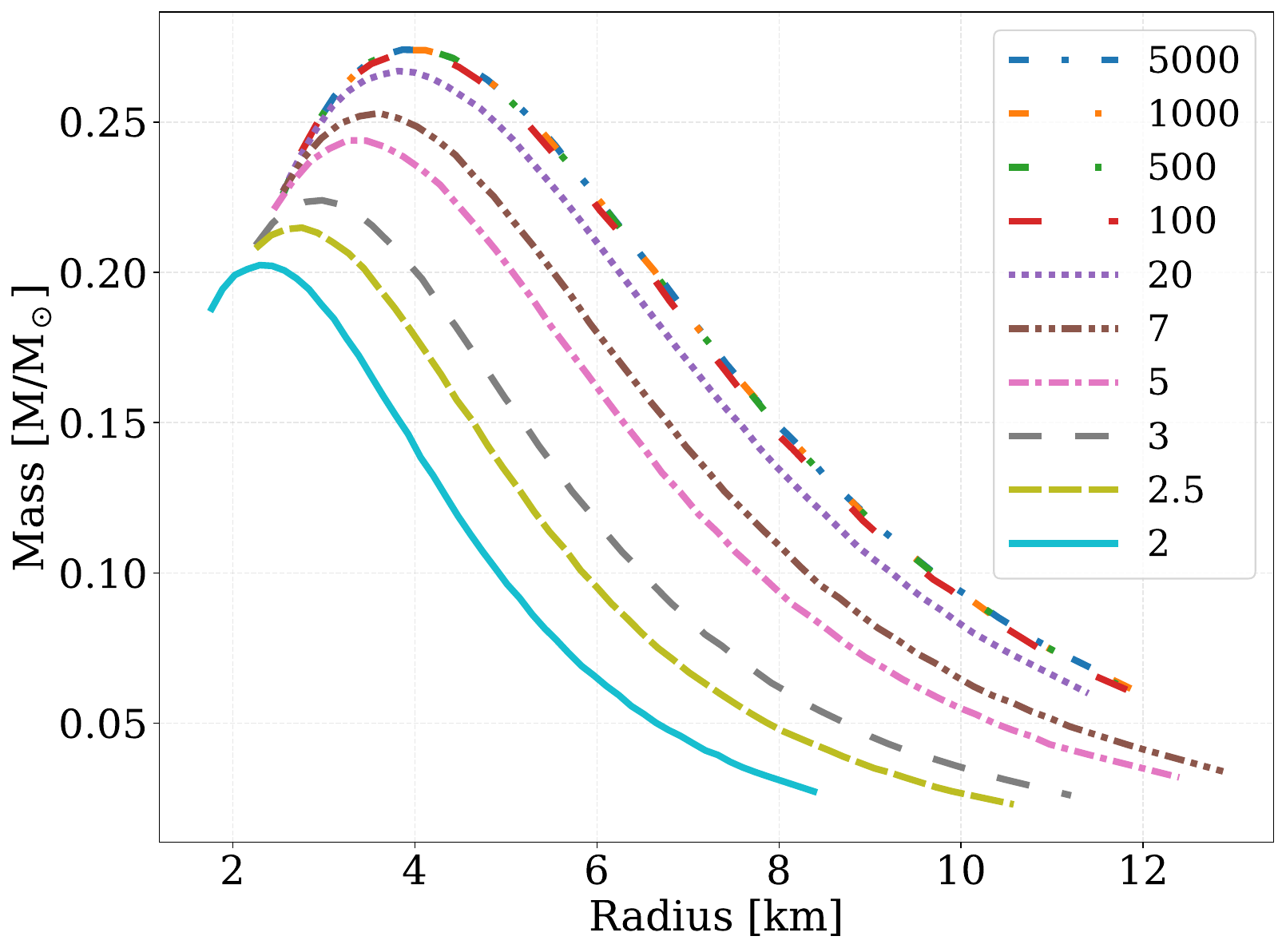}
    \caption{\textit{Left}: Pressure vs. energy density for DSR at supra-nuclear saturation densities. Nuclear saturation density ($n_0$) is $0.15 ~\rm{nucleons}~\rm{fm}^{-3}$, energy density $\sim 0.71~\rm{fm}^{-4}$. \textit{Right}: Mass-Radius relations for neutron stars obtained for the DSR. Curves in the colors blue, orange, green, red, purple, brown, pink, gray, lime, and aqua are for $\kappa = 5000, 1000, 500, 100, 20, 7, 5, 3, 2.5, \text{and}~2~ \mathrm{GeV}$ respectively.}
    \label{fig:ns-MR}
\end{figure}

The modified dispersion relation obtained in Sec.~\ref{sec:T=0} for $T\rightarrow 0$ describes the thermodynamic relationship for a Fermi gas of neutrons and can be applied to dense neutron star matter. The DSR produces a unique relationship in nuclear behavior in response to increase in density. We apply the relation obtained between $P$ and $n$ in Eqs.~\eqref{eq:t_0_P} and \eqref{eq:t_0_E} for $\kappa=2,2.5,7,20,100,500,1000~\rm{and}~5000~\rm{\mathrm{GeV}}$ and solve the coupled differential equations --- Tolman--Oppenheimer--Volkoff (TOV) equation and total mass equation (Eq. \eqref{eq:total-mass}) for hydrostatic neutron stars supported \cite{Oppenheimer:1939ne, Tolman:1939jz}. The equations of state produced by the DSR are causal ( $c_s< 1$) at all densities with the highest sound speed in each $\kappa$ model $c_{s,\mathrm{max}}\in[0.54, 0.58]$. Thermodynamic stability is also retained throughout the fluid due to $\mathrm{d} P/\mathrm{d}\epsilon\geq0$ at all densities.

In Fig.~\ref{fig:ns-MR} left panel, the equation of state pressure vs. total energy density for a Fermi gas of neutrons based on Eqs. \eqref{eq:t_0_P} and \eqref{eq:t_0_E} from the DSR is shown. Dynamical stability test using the Harrison–Zel’dovich–Novikov criterion, $\mathrm{d} M/\mathrm{d}\rho_c \geq0$ (where $\rho_c$ is the density at the core), \cite{1965gtgc.book.....H, 1971reas.book.....Z} shows that the branch of neutron stars from low mass-large radii to the highest mass is stable, whereas the stars after the turning point are solutions that are in an unstable equilibrium. The DSR produces lower pressures than nucleonic pressure-density relationships, and the pressure decreases substantially with smaller values of $\kappa$ to result in \textit{soft} equations of state. The nucleonic equation of state SLy4 \cite{1998NuPhA.635..231C} attains a pressure of $2.1~\rm{fm}^{-4}$ at an energy density of $10~\rm{fm}^{-4}$, i.e. a higher pressure than the DSR. $\kappa=2~\rm{\mathrm{GeV}}$ is on the lower limit of feasible $\kappa$ because $m=m_n = 0.939~\rm{\mathrm{GeV}}$ for neutron star matter. Fig.~\ref{fig:ns-MR} right panel shows the neutron star mass-radius curves for stable non-rotating neutron stars obtained by solving the TOV equation. The mass-radius curve shows the family of neutron stars in hydrostatic equilibrium supported by the specific equation of state (EoS), where each point represents a neutron star with a unique density configuration. Values of $\kappa$ that are a few times that of $m_n$ show lower radii for same masses, showing that the inclusion of a strong modification that leads to substantially \textit{soft} equations of state also constrict the material into a small volume to form compact neutron stars. The heaviest neutron star supported by the DSR model is lighter than observed neutron star masses, HESS J1731-347 with $M \approx (0.7\pm 0.2)  M_\odot$ and radius $R \approx (10.4\pm 0.8)~\rm{km}$ \cite{HESS}. 

\section{Conclusion and Outlook}
\label{sec:conclusion}
In this work we have investigated the thermodynamic properties of a relativistic Fermi gas governed by a modified dispersion relation arising in the Magueijo–Smolin (MS) model \cite{Magueijo_2002} of Doubly Special Relativity (DSR), characterized by the presence of an observer independent ultraviolet energy scale $\kappa$. Our analysis was carried out in two physically distinct regimes: the near-degenerate low-temperature limit $T \rightarrow 0$, which is relevant for compact astrophysical objects like white dwarfs and neutron stars, and the high-temperature regime $m \ll T \ll \kappa$. 

In the low-temperature regime $T\rightarrow 0$, we employ a Sommerfeld expansion to obtain analytic expressions for the thermodynamic quantities of the deformed Fermi gas. The resulting expressions are formally similar to those obtained in the special relativistic case at low- temperature, but with a modified density of states determined by the DSR dispersion relation. We plot the thermodynamic quantities as functions of the number density for neutrons in Fig.~\ref{fig:low-temp}. From these results, we observe that values of the deformation scale in the range $0.939\,\mathrm{\mathrm{GeV}} < \kappa \lesssim 10\,\mathrm{\mathrm{GeV}}$ lead to appreciable deviations from the standard special relativistic behavior. We further use the equation of state obtained in this regime as input to the Tolman–Oppenheimer–Volkoff (TOV) equations to construct hydrostatic neutron star configurations.   

In the high-temperature regime, the thermodynamic quantities were evaluated numerically from the exact grand canonical potential $\Omega$. In this limit, both the deformed and the special relativistic cases follow the standard ideal gas behavior in their respective frameworks, and the corresponding thermodynamic functions coincide within numerical accuracy in the temperature range considered \cite{das2024ideal}. This confirms that the high-temperature limit of the Fermi gas is governed by ideal gas thermodynamics and that quantum degeneracy effects are negligible as long as $T \ll \kappa$.

To understand the impact of DSR in physical systems, we utilize our low-temperature results to solve for static non-rotating white dwarf and neutron star using structure differential relationships. We conduct this for a range of DSR parameter, $\kappa$, taking it from highly modified to mildly modified w.r.t. SR as $\kappa$ goes from $\kappa\gtrsim m_b$ to $\kappa>>m_b$. For both white dwarfs and neutron stars we find a strong dependence of internal structure on $\kappa$. For white dwarfs with Helium core we find a strong dependence on $\kappa$, whereas for heavier cores we notice degeneracies in solutions that strongly support a range of DSRs interchangeable with a range of core component, $m_b$. The equation of state for cold degenerate neutrons is \textit{softened} by the DSR which produces small ($R<10~\rm{km}$) and light ($M<0.3~\rm{M_\odot}$) neutron stars. While these are smaller than known neutron star observations these solutions may point towards a separate branch of exotic stars and towards hybrid stars with mild-DSR and non-DSR regions determined by a density dependent $\kappa$. In future work, the dynamical evolution of white dwarf mergers and neutron star mergers with DSR should also be studied as those would highlight the impacts of the modified density profile.

Codes to solve for the Fermi gas thermodynamic quantities are available in the form of Mathematica notebooks on \url{https://github.com/AtulKedia93/Fermi_gas_DSR}

\section*{Appendix}
\appendix

\section{Derivation of density of states}
The density of states is defined as the following,
\begin{align}
    \rho (E)= \frac{4\pi V g}{h^3} p^2 \frac{dp}{dE} \label{eq:rho_original}
\end{align}
For the dispersion relation
\begin{align}
    &E^2-p^2=m^2\left(1-\frac{E}{\kappa}\right)^2\,,
\end{align}
we get
\begin{align}
    &\text{d}p
    =\bigg[
    E^2
    -m^2\left(1-\frac{E}{\kappa}\right)^{2}
    \bigg]^{-\frac{1}{2}}
    \bigg[
    E
    +\frac{m^2}{\kappa}\left(1-\frac{E}{\kappa}\right)
    \bigg]
    \text{d}E\,. \label{eq:dp}
\end{align}
Using Eq. \eqref{eq:dp} in Eq. \eqref{eq:rho_original} we obtain:
\begin{align}
    \rho(E)
    =\frac{4\pi V g}{h^3}
    \Bigg[E^2-m^2\left(1-\frac{E}{\kappa}\right)^2\Bigg]^{1/2}
    \left[E
    +\frac{m^2}{\kappa}\left(1-\frac{E}{\kappa}\right)\right]\,.\label{eq:rho_final}
\end{align}

\section{Sommerfeld expansion of pressure}
\label{app:pressure_expansion}

In this section we show the Sommerfeld expansion of pressure explicitly. Starting from Eq. \eqref{eq:ibp}:
\begin{equation}
\label{eq:appibp}
    P = \frac{g}{6\pi^2}\int_{m_{0}}^{\kappa} dE \frac{\overbrace{\bigg[E^{2}-m^{2}\bigg(1-\frac{E}{\kappa}\bigg)^{2}\bigg]^{3/2}}^{H(E)}}{\bigg[1+\exp[\beta(E-\mu)]\bigg]} ~~~.
\end{equation}
Let's call $H(E)=\bigg[E^{2}-m^{2}\bigg(1-\frac{E}{\kappa}\bigg)^{2}\bigg]^{3/2}$. We now define $K(E)$ as

\begin{align}
    K(E)=
    \int_{m_0}^EdE^\prime H(E^\prime)
    \implies
    H(E)=
    \frac{dK(E)}{dE}\,.
\end{align}
Upon substitution of $H(E)$ in Eq. \eqref{eq:appibp}, the expression of pressure now becomes the following.
\begin{align}
\label{eq:Plb}
     P =
     \frac{g}{6\pi^2}
     \underbrace{\int_{m_{0}}^{\kappa} dE 
     \frac{dK(E)}{dE}n(E)}_{\langle H\rangle}
\end{align}
with $n(E)$ being the usual Fermi-Dirac distribution. Let's now focus on the integration. We name the integration as $\langle H\rangle$. Upon doing integration by parts we get
\begin{align}
\label{eq:<H>}
    \langle H\rangle=
    -\int_{m_0}^\kappa dE K(E)\frac{dn(E)}{dE}\,.
\end{align}
The surface term vanishes because $H(m_0)=0$ and $n(\kappa)=0$. Now we Taylor expand $K(E)$ around $E=\mu$.
\begin{align}
    K(E)
    =K(\mu)
    +(E-\mu)K'(\mu)
    +\frac{(E-\mu)^2}{2}K''(\mu)
\end{align}
Substituting this back in Eq. \eqref{eq:<H>} gives
\begin{align}
\label{eq:Hexp}
    \langle H\rangle=
    -\int_{m_0}^\kappa dE \left[K(\mu)
    +(E-\mu)K'(\mu)
    +\frac{(E-\mu)^2}{2}K''(\mu)\right]\frac{dn(E)}{dE}\,.
\end{align}
We look at each term separately. The first term:
\begin{align}
    -\int_{m_0}^\kappa dE \, K(\mu)\frac{dn(E)}{dE}
    =&(n(m_0)-n(\kappa))K(\mu)\nonumber\\
    =&\int_{m_0}^\mu dEH(E)
\end{align}
The last step follows from $n(\kappa)=0$ and $n(m_0)=1$. This is the result for $T=0$. Now let us at the second term of Eq. \eqref{eq:Hexp}.
\begin{align}
\label{eq:Hexp2}
    -&\int_{m_0}^\kappa dE
    (E-\mu)K'(\mu)\frac{dn(E)}{dE}\nonumber
    =\int_{m_0}^\kappa dE(E-\mu)K'(\mu)\beta 
    \frac{e^{\beta(E-\mu)}}{(1+e^{\beta(E-\mu)})^2}\nonumber\\
    =&H(\mu)\beta
    \bigg[\int_{m_0}^\mu dE(E-\mu) 
    \frac{e^{\beta(E-\mu)}}{(1+e^{\beta(E-\mu)})^2}
    +\int_{\mu}^\kappa dE(E-\mu) 
    \frac{e^{\beta(E-\mu)}}{(1+e^{\beta(E-\mu)})^2}\bigg]\nonumber\\
    =&H(\mu)\beta\frac{1}{2\beta^2}
    \bigg[
    -\frac{2 e^{m_0\beta}m_0\beta}{e^{m_0\beta}+e^{\beta\mu}}
    +\beta\mu-\ln{4}
    -2\beta\mu
    +2\ln(e^{m_0\beta}+e^{\beta\mu})
    +\beta\mu\tanh(\frac{1}{2}\beta(m_0-\mu))\nonumber\\
    &+\frac{2 e^{\kappa\beta}\kappa\beta}{e^{\kappa\beta}+e^{\beta\mu}}
    -\beta\mu+\ln{4}
    +2\beta\mu
    -2\ln(e^{\kappa\beta}+e^{\beta\mu})
    -\beta\mu\tanh(\frac{1}{2}\beta(\kappa-\mu))
    \bigg]
\end{align}
We arrive at the last step by carrying out the integration with respect to $E$.
\begin{align}
    \frac{2 e^{m_0\beta}m_0\beta}{e^{m_0\beta}+e^{\beta\mu}}
    =\frac{2 e^{\beta(m_0-\mu)}m_0\beta}{e^{\beta(m_0-\mu)}+1}
    =0
\end{align}
This is because $m_0-\mu<0$ which leads to $e^{\beta(m_0-\mu)}\to e^{-\infty}$. Although it is multiplied with with a factor of $\beta$, the exponential part is going to dominate. Hence the whole term approaches 0. Similarly,
\begin{align}
    \frac{2 e^{\kappa\beta}\kappa\beta}{e^{\kappa\beta}+e^{\beta\mu}}
    =\frac{2 e^{\beta(\kappa-\mu)}\kappa\beta}{e^{\beta(\kappa-\mu)}+1}=2\kappa\beta
\end{align}
This is due to the fact that $e^{\beta(\kappa-\mu)}+1=e^{\beta(\kappa-\mu)}$ as $\kappa-\mu>0$ leading to $e^{\beta(\kappa-\mu)}\to\infty$. Hence, 
\begin{align}
    2\ln(e^{m_0\beta}+e^{\beta\mu})
    =2\beta\mu \hspace{2em} \rm{and}\hspace{2em} 
    2\ln(e^{\kappa\beta}+e^{\beta\mu})
    =~2\beta\kappa ~~~.
\end{align}
Now looking at the hyperbolic functions:
\begin{align}
    \beta\mu\tanh(\frac{1}{2}\beta(m_0-\mu))
    =&\beta\mu
    \frac{e^{\frac{\beta(m_0-\mu)}{2}}-e^{-\frac{\beta(m_0-\mu)}{2}}}{e^{\frac{\beta(m_0-\mu)}{2}}+e^{-\frac{\beta(m_0-\mu)}{2}}}
    =-\beta\mu \hspace{2em} \rm{and}\hspace{2em} \beta\mu\tanh(\frac{1}{2}\beta(\kappa-\mu))
    =\beta\mu
\end{align}
Hence, the final result for Eq. \eqref{eq:Hexp2} is
\begin{align}
    -\int_{m_0}^\kappa dE
    (E-\mu)K'(\mu)\frac{dn(E)}{dE} =&H(\mu)\frac{1}{2\beta}
    \bigg[
    \beta\mu-\ln{4}
    -2\beta\mu
    +2\beta\mu
    -\beta\mu
    +2\beta\kappa
    -\beta\mu+\ln{4}
    +2\beta\mu
    -2\beta\kappa
    -\beta\mu
    \bigg]\nonumber\\
    =&~0~~~.
\end{align}
Now we look at the third term of Eq. \eqref{eq:Hexp}.
\begin{align}
\label{eq:Hexp3}
    -&\int_{m_0}^\kappa
    \frac{(E-\mu)^2}{2}K''(\mu)\frac{dn(E)}{dE} = K''(\mu)
    \frac{\beta}{2}
    \bigg[
    \int_{m_0}^\mu dE
    (E-\mu)^2 
    \frac{e^{\beta(E-\mu)}}{(1+e^{\beta(E-\mu)})^2}
    +\int_{\mu}^\kappa dE
    (E-\mu)^2 
    \frac{e^{\beta(E-\mu)}}{(1+e^{\beta(E-\mu)})^2}
    \bigg]\nonumber\\
    =&K''(\mu)
    \frac{\beta}{2}
    \frac{1}{\beta^3}
    \bigg[
    -\beta(\kappa-\mu)
    \left(-\frac{e^{\kappa\beta}\beta(\kappa-\mu)}{e^{\kappa\beta}+e^{\mu\beta}}
    +2\ln(1+e^{\beta(\kappa-\mu)})\right)\nonumber\\
    &+\beta(m_0-\mu)
    \left(-\frac{e^{m_0\beta}\beta(m_0-\mu)}{e^{m_0\beta}+e^{\mu\beta}}
    +2\ln(1+e^{\beta(m_0-\mu)})\right)
    -2~\text{Li}_2(-e^{\beta(\kappa-\mu)})
    +2~\text{Li}_2(-e^{\beta(m_0-\mu)})
    \bigg] ~~,
\end{align}
where $\text{Li}_2$ is polylogarithm of order 2. We take a similar approach as taken for the second term of the expansion.
\begin{align}
    \frac{e^{\kappa\beta}\beta(\kappa-\mu)}{e^{\kappa\beta}+e^{\mu\beta}}
    &= \frac{e^{\beta(\kappa-\mu)}\beta(\kappa-\mu)}{e^{\beta(\kappa-\mu)}+1}
    =\beta(\kappa-\mu) \\
    \frac{e^{m_0\beta}\beta(m_0-\mu)}{e^{m_0\beta}+e^{\mu\beta}}
    &=\frac{e^{\beta(m_0-\mu)}\beta(m_0-\mu)}{e^{\beta(m_0-\mu)}+1}=0 \\
    2\ln(1+e^{\beta(\kappa-\mu)})
    &=2\beta(\kappa-\mu) \\
    2\ln(1+e^{\beta(m_0-\mu)})&=0
\end{align}
We asymptotically expand polylogarithmic function. For any very large $x$, $\text{Li}_2(-e^{x})=-\frac{\pi^2}{6}-\frac{x^2}{2}$. So we can write the following.
\begin{align}
    2~\text{Li}_2(-e^{\beta(\kappa-\mu)})
    =-\frac{\pi^2}{3}-\beta^2(\kappa-\mu)^2
    \hspace{2em} \rm{and}\hspace{2em}
    2~\text{Li}_2(-e^{\beta(m_0-\mu)})
    =0
\end{align}
This is because $\text{Li}_2(0)=0$. Hence, Eq. \eqref{eq:Hexp3} finally yields
\begin{align}
    -\int_{m_0}^\kappa 
    \frac{(E-\mu)^2}{2}K''(\mu)\frac{dn(E)}{dE} =& K''(\mu)
    \frac{\beta}{2}
    \frac{1}{\beta^3}
    \bigg[
    -\beta^2(\kappa-\mu)^2
    +\frac{\pi^2}{3}+\beta^2(\kappa-\mu)^2
    \bigg]\nonumber\\
    =&H'(\mu)\frac{1}{\beta^2}\frac{\pi^2}{6}\,~~~.
\end{align}
Therefore,
\begin{align}
\label{eq:<H>f}
    \langle H\rangle
    =&\int_{m_0}^\mu dEH(E)
    +H'(\mu)\frac{1}{\beta^2}\frac{\pi^2}{6}\,.
\end{align}
Substituting Eq. \eqref{eq:<H>f} in Eq. \eqref{eq:Plb} culminates to the following expression for pressure (same as Eq. \eqref{eq:t_0_P})
\begin{align}
    P &= \frac{g}{6\pi^2} \Bigg\{\int_{m_0}^\mu \text{d} E \bigg[ E^{2} - m^{2} \bigg( 1 - \frac{E}{\kappa} \bigg)^{2} \bigg]^{3/2} + \frac{\pi^2}{4} \frac{1}{\beta^2} \Bigg[ \left[ 2 \mu + \frac{2 m^2}{\kappa} \left( 1 - \frac{\mu}{\kappa} \right)  \right] \bigg[ \mu^2 - m^2 \left( 1 - \frac{\mu}{\kappa} \right)^2 \bigg]^{1/2} \Bigg]
    +\mathcal{O}\left(\frac{1}{\beta^3}\right)\Bigg\}.
\end{align}
\section{Sommerfeld chemical potential}
The number density is given by:
\begin{align}
    n(T,\mu)
    =\int_{m_0}^\kappa dE \rho(E)n(E)\,,
\end{align}
where $\rho(E)$ is the density of states for our theory. The Sommerfeld expansion of the number density is as follows:
\begin{align}
    n(T,\mu)
    =&\int_{m_0}^\mu dE\rho(E)
    +\rho'(\mu)\frac{1}{\beta^2}\frac{\pi^2}{6}\nonumber\\
    =&\int_{m_0}^{\epsilon_f} dE\rho(E)
    +\int_{\epsilon_f}^\mu dE\rho(E)
    +\rho'(\mu)\frac{1}{\beta^2}\frac{\pi^2}{6}\,,
\end{align}
with $\epsilon_f$ being the Fermi energy. The first term is the number density at T=0. For the number density to remain constant throughout, we need
\begin{align}
    0=\int_{\epsilon_f}^\mu dE\rho(E)
    +\rho'(\mu)\frac{1}{\beta^2}\frac{\pi^2}{6}\,.
\end{align}
Assuming $\rho(E)$ to not change around $\epsilon_f$ and also assuming $\rho'(\epsilon_f)=\rho'(\mu)$,
\begin{align}
    0=&(\mu-\epsilon_f)+\frac{\rho'(\epsilon_f)}{\rho(\epsilon_f)}\frac{1}{\beta^2}\frac{\pi^2}{6}\nonumber\\
    \mu
    =&\epsilon_f-\frac{\rho'(\epsilon_f)}{\rho(\epsilon_f)}\frac{1}{\beta^2}\frac{\pi^2}{6}\,.\label{eq:mu}
\end{align}
The Fermi energy for our dispersion relation is
\begin{align}
    \epsilon_f
    =\frac{1}{1-\frac{m^2}{\kappa^2}}
    \Bigg[
    -\frac{m^2}{\kappa}
    +\sqrt{m^2+\bigg(1-\frac{m^2}{\kappa^2}\bigg)\bigg(\frac{3h^3}{4\pi g}n\bigg)^{2/3}}\Bigg]\,.\label{eq:fermienergy}
\end{align}
Using Eq. \eqref{eq:rho_final} and Eq. \eqref{eq:fermienergy} in Eq. \eqref{eq:mu} gives:
\begin{align}
\mu &=
\frac{
k\left(
-2m^2
+
k\sqrt{
4m^2
+
\left(1-\frac{m^2}{k^2}\right)
\left(\frac{h^3 n}{g}\right)^{2/3}
\left(\frac{6}{\pi}\right)^{2/3}
}
\right)
}
{2\left(k^2-m^2\right)}\nonumber\\
&+
\frac{
2^{1/3}
\left[
6^{2/3}m^2\left(\frac{h^3 n}{g}\right)^{2/3}
-
k^2\left(
6^{2/3}\left(\frac{h^3 n}{g}\right)^{2/3}
+
2m^2\pi^{2/3}
\right)
\right]
\pi^{7/3}
}
{
3\,3^{2/3}\,
k^2
\left(\frac{h^3 n}{g}\right)^{2/3}
\sqrt{
6^{2/3}\left(\frac{h^3 n}{g}\right)^{2/3}
-
\frac{
6^{2/3}m^2\left(\frac{h^3 n}{g}\right)^{2/3}
}{k^2}
+
4m^2\pi^{2/3}
}
\,\beta^2
}.
\label{eq:mu_final}
\end{align}
Using Eq. \eqref{eq:fermi_mom} and Eq. \eqref{eq:Lambda} after some algebra one can reduce Eq. \eqref{eq:mu_final} to:
\begin{align}
    \mu=\frac{\kappa(\kappa \Lambda -m^2)}{\kappa^2 - m^2} - \frac{\pi^{2} \left[m^2 + 2p_F^2 \left(1-\frac{m^2}{\kappa^2}\right)\right]}{6 \beta^2 p_F^2 \Lambda}\,. \label{eq:mu_reduced}
\end{align}
As $\kappa\to\inf$, $\Lambda\to \sqrt{m^2+p_F^2}=\epsilon_{f_{\rm SR}}$, which is the fermi energy of a relativistic fermi gas. The chemical potential in Eq. \eqref{eq:mu_reduced} reduces to:
\begin{align}
    \lim_{\kappa\to\infty} \mu = \epsilon_{f_{\rm SR}} - \frac{\pi^{2} \left[m^2 + 2p_F^2 \left(1-\frac{m^2}{\kappa^2}\right)\right]}{6 \beta^2 p_F^2 \epsilon_{f_{\rm SR}}}\,.
\end{align}
From Eq. \eqref{eq:mu} we have:
\begin{align}
    \mu
    =&\epsilon_f-\frac{\rho'(\epsilon_f)}{\rho(\epsilon_f)}\frac{1}{\beta^2}\frac{\pi^2}{6}\,.
\end{align}
We can see that at $T=0$, i.e., $\beta\to\infty$, we get $\mu=\epsilon_{f_{\rm SR}}$.
\section{Sommerfeld expansion of entropy}
\begin{align}
    S=&
    -\left(\frac{\partial \Omega}{\partial 
    T}\right)_{V,\,\mu}\nonumber\\
    =&\frac{\partial}{\partial T}\left(\frac{1}{\beta}\ln{Z}\right)\nonumber\\
    =&\frac{\partial}{\partial T}
    \Bigg(
    \frac{4\pi g V}{\beta h^3}
    \int_0^\kappa \text{d}p p^2\ln{(1+e^{-\beta(E-\mu)})}
    \Bigg)\nonumber\\
    =&\frac{4\pi g V}{h^3}
    \left(\frac{1}{k_BT^2}\right)
    \Bigg[
    \frac{1}{\beta^2}C_1
    +\frac{1}{\beta}\int_0^\kappa\text{d}p\frac{p^3}{3}
    \frac{1}{e^{\beta(E-\mu)}+1}\frac{\text{d}E}{\text{d}p}
    +\frac{1}{\beta}\int_0^\kappa\text{d}p p^2
    \frac{E-\mu}{e^{\beta(E-\mu)}+1}
    \Bigg]
\end{align}

$C_1=0$.

\begin{align}
    S=&
    \frac{4\pi g V}{h^3}
    \left(\frac{1}{k_BT^2}\right)
    \int_{m_0}^\kappa \frac{\text{d}E}{1+e^{\beta(E-\mu)}}
    \Bigg[
    \frac{1}{3\beta}
    \left(E^2-m^2\bigg(1-\frac{E}{\kappa}\bigg)^2\right)^{3/2}\nonumber\\
    &+\frac{1}{\beta}\left(E^2-m^2\bigg(1-\frac{E}{\kappa}\bigg)^2\right)^{1/2}
    \left(E+\frac{m^2}{\kappa}\bigg(1-\frac{E}{\kappa}\bigg)\right)
    (E-\mu- \beta \frac{\partial \mu}{\partial \beta})
    \Bigg]
\end{align}

For $T\to0$, the entropy can be effectively written as

\begin{align}
    S=&\frac{4\pi gV}{h^3}
    \left(k_B\beta^2\right)
    \bigg\{\int_{m0}^\mu dE
    \bigg[\frac{1}{3\beta} \left(E^2-m^2\bigg(1-\frac{E}{\kappa}\bigg)^2\right)^{3/2}
    +\frac{1}{\beta}\left(E^2-m^2\bigg(1-\frac{E}{\kappa}\bigg)^2\right)^{1/2}
    \left(E+\frac{m^2}{\kappa}\bigg(1-\frac{E}{\kappa}\bigg)\right)\nonumber\\
    &(E-\mu- \beta \frac{\partial \mu}{\partial \beta})\bigg]+\frac{1}{\beta^2}\frac{\pi^2}{6}
    \frac{2}{\kappa^2\beta}(\kappa m^2+\kappa^2\mu-m^2\mu)
    \bigg(
    \mu^2-\frac{m^2}{\kappa^2}(\kappa-\mu)^2
    \bigg)^{1/2}
    \bigg\} \label{eq:entropy1}
\end{align}
For matching it to the order of temperature in pressure we need atleast 2 more terms in the expansion of $\langle H\rangle$. 

\begin{align}
    &-\int_{m_0}^\kappa dE
    \frac{(E-\mu)^3}{3!}K^{\prime\prime\prime}(\mu)
    \frac{dn(E)}{dE}\nonumber\\
    =&K^{\prime\prime\prime}(\mu)
    \frac{\beta}{3!}
    \bigg[
    \int_{m_0}^\mu dE
    (E-\mu)^3
    \frac{e^{\beta(E-\mu)}}{(e^{\beta(E-\mu)}+1)^2}
    +\int_{\mu}^\kappa dE
    (E-\mu)^3
    \frac{e^{\beta(E-\mu)}}{(e^{\beta(E-\mu)}+1)^2}
    \bigg]\nonumber\\
    =&K^{\prime\prime\prime}(\mu)
    \frac{\beta}{3!}
    \bigg[
    -\frac{(m_0-\mu)^3}{\beta}
    +\frac{(m_0-\mu)^3}{\beta(1+e^{\beta(m_0-\mu)})}
    +\frac{3(m_0-\mu)^2\ln(1+e^{\beta(m_0-\mu)})}{\beta^2}
    +\frac{6(m_0-\mu)Li_2(-e^{\beta(m_0-\mu)})}{\beta^3}\nonumber\\
    &-\frac{6Li_3(-e^{\beta(m_0-\mu)})}{\beta^4}
    -\frac{9}{2}\frac{\zeta(3)}{\beta^4}
    +\frac{(\kappa-\mu)^3}{\beta}
    -\frac{(\kappa-\mu)^3}{\beta(1+e^{\beta(\kappa-\mu)})}
    -\frac{3(\kappa-\mu)^2\ln(1+e^{\beta(\kappa-\mu)})}{\beta^2}
    -\frac{6(\kappa-\mu)Li_2(-e^{\beta(\kappa-\mu)})}{\beta^3}\nonumber\\
    &+\frac{6Li_3(-e^{\beta(\kappa-\mu)})}{\beta^4}
    +\frac{9}{2}\frac{\zeta(3)}{\beta^4}
    \bigg]\nonumber\\
    =&K^{\prime\prime\prime}(\mu)
    \frac{\beta}{3!}
    \bigg[
    -\frac{(m_0-\mu)^3}{\beta}
    +\frac{(m_0-\mu)^3}{\beta}
    +\frac{3(m_0-\mu)^2\ln(1)}{\beta^2}
    +\frac{6(m_0-\mu)Li_2(0)}{\beta^3}\nonumber\\
    &-\frac{6Li_3(0)}{\beta^4}
    +\frac{(\kappa-\mu)^3}{\beta}
    -\frac{(\kappa-\mu)^3}{\beta(e^{\beta(\kappa-\mu)})}
    -\frac{3(\kappa-\mu)^2\ln(e^{\beta(\kappa-\mu)})}{\beta^2}
    -\frac{6(\kappa-\mu)Li_2(-e^{\beta(\kappa-\mu)})}{\beta^3}
    +\frac{6Li_3(-e^{\beta(\kappa-\mu)})}{\beta^4}
    \bigg]\nonumber\\
    =&K^{\prime\prime\prime}(\mu)
    \frac{\beta}{3!}
    \bigg[
    +\frac{(\kappa-\mu)^3}{\beta}
    -\frac{3(\kappa-\mu)^3}{\beta}
    +\frac{6(\kappa-\mu)}{\beta^3}\frac{\pi^2}{6}
    +\frac{6(\kappa-\mu)}{\beta^3}\frac{\beta^2(\kappa-\mu)^2}{2}
    -\frac{6}{\beta^4}\frac{\beta^3(\kappa-\mu)^3}{6}
    -\frac{6}{\beta^4}\frac{\pi^2\beta(\kappa-\mu)}{6}
    \bigg]\nonumber\\
    =&0
\end{align}

where, upon asymptotic expansion,

\begin{align}
    Li_3(-e^x)=-\frac{x^3}{6}-\frac{\pi^2}{6}x
\end{align}

And also,

\begin{align}
    &-\int_{m_0}^\kappa dE
    \frac{(E-\mu)^4}{4!}K^{\prime\prime\prime\prime}(\mu)
    \frac{dn(E)}{dE}\nonumber\\
    =&K^{\prime\prime\prime\prime}(\mu)
    \frac{\beta}{4!}
    \bigg[
    \int_{m_0}^\mu dE
    (E-\mu)^4
    \frac{e^{\beta(E-\mu)}}{(e^{\beta(E-\mu)}+1)^2}
    +\int_{\mu}^\kappa dE
    (E-\mu)^4
    \frac{e^{\beta(E-\mu)}}{(e^{\beta(E-\mu)}+1)^2}
    \bigg]\nonumber\\
    =&K^{\prime\prime\prime\prime}(\mu)
    \frac{\beta}{4!}
    \bigg[
    \frac{1}{30e^{\beta m_0}(1+e^{\beta(\mu-m_0)})\beta^5}
    \bigg(
    7\pi^4e^{\beta m_0}(1+e^{\beta(\mu-m_0)})
    -30 e^{\beta m_0}\beta^4 (\mu-m_0)^4\nonumber\\
    &+120e^{\beta m_0} \beta^3 (\mu-m_0)^3\ln(1+e^{\beta(m_0-\mu)})(1+e^{\beta(\mu-m_0)})
    +360 e^{\beta m_0}(1+e^{\beta(\mu-m_0)})\beta^2(m_0-\mu)^2 Li_2(-e^{\beta(m_0-\mu)})\nonumber\\
    &-720 e^{\beta m_0}(1+e^{\beta(\mu-m_0)})\beta(m_0-\mu) Li_3(-e^{\beta(m_0-\mu)})
    +720 e^{\beta m_0}(1+e^{\beta(\mu-m_0)})Li_4(-e^{\beta(m_0-\mu)})
    \bigg)\nonumber\\
    &-\frac{1}{30e^{\beta \kappa}(1+e^{\beta(\mu-\kappa)})\beta^5}
    \bigg(7\pi^4e^{\beta \kappa}(1+e^{\beta(\mu-\kappa)})
    -30 e^{\beta \kappa}\beta^4 (\mu-\kappa)^4\nonumber\\
    &+120e^{\beta \kappa} \beta^3 (\mu-\kappa)^3\ln(1+e^{\beta(\kappa-\mu)})(1+e^{\beta(\mu-\kappa)})
    +360 e^{\beta \kappa}(1+e^{\beta(\mu-\kappa)})\beta^2(\kappa-\mu)^2 Li_2(-e^{\beta(\kappa-\mu)})\nonumber\\
    &-720 e^{\beta\kappa}(1+e^{\beta(\mu-\kappa)})\beta(\kappa-\mu) Li_3(-e^{\beta(\kappa-\mu)})
    +720 e^{\beta\kappa}(1+e^{\beta(\mu-\kappa)})Li_4(-e^{\beta(\kappa-\mu)})
    \bigg)
    \bigg]\nonumber\\
    =&K^{\prime\prime\prime\prime}(\mu)
    \frac{\beta}{4!}
    \bigg[
    \frac{1}{30e^{\beta m_0}(e^{\beta(\mu-m_0)})\beta^5}
    \bigg(
    7\pi^4e^{\beta m_0}(e^{\beta(\mu-m_0)})
    -30 e^{\beta m_0}\beta^4 (\mu-m_0)^4
    \bigg)\nonumber\\
    &-\frac{1}{30e^{\beta \kappa}\beta^5}
    \bigg(7\pi^4e^{\beta \kappa}(1+e^{\beta(\mu-\kappa)})
    -30 e^{\beta \kappa}\beta^4 (\mu-\kappa)^4\nonumber\\
    &-120e^{\beta \kappa} \beta^4 (\kappa-\mu)^4
    -360 e^{\beta \kappa}\beta^2(\kappa-\mu)^2 \frac{\pi^2}{6}
    -360 e^{\beta \kappa}\beta^2(\kappa-\mu)^2 \frac{(\beta(\kappa-\mu))^2}{2}\nonumber\\
    &+720 e^{\beta\kappa}(\beta(\kappa-\mu))^2 \frac{\pi^2}{6}
    +720 e^{\beta\kappa} \frac{(\beta(\kappa-\mu))^4}{6}
    -720 e^{\beta\kappa}\frac{7\pi^4}{360}
    -720 e^{\beta\kappa}(\beta(\kappa-\mu))^2\frac{\pi^2}{12}
    -720 e^{\beta\kappa}\frac{(\beta(\kappa-\mu))^4}{24}
    \bigg)
    \bigg]\nonumber\\
    =&K^{\prime\prime\prime\prime}(\mu)
    \frac{\beta}{4!}
    \frac{14 \pi^4}{30\beta^5}
\end{align}

 where, upon asymptotic expansion,

\begin{align}
    Li_4(-e^x)=-\frac{7\pi^4}{360}
    -\frac{1}{12}\pi^2x^2-\frac{1}{24}x^4\,.
\end{align}

We don't make use of these extra terms as the term of the order $T^2$ is 0.

We can reduce the expression of entropy in Eq.\eqref{eq:entropy1} by the differentiating by parts. Eq. \eqref{eq:entropy1} reads:
\begin{align}
    S=&\frac{4\pi gV}{h^3}
    \left(k_B\beta^2\right)
    \bigg\{\int_{m0}^\mu dE
    \bigg[\frac{1}{3\beta} \left(E^2-m^2\bigg(1-\frac{E}{\kappa}\bigg)^2\right)^{3/2}
    +\frac{1}{3\beta}\frac{d}{dE}\left(E^2-m^2\bigg(1-\frac{E}{\kappa}\bigg)^2\right)^{3/2}(E-\mu)\bigg]\nonumber\\
    &+\frac{1}{\beta^2}\frac{\pi^2}{6}
    \frac{2}{\kappa^2\beta}(\kappa m^2+\kappa^2\mu-m^2\mu)
    \bigg(
    \mu^2-\frac{m^2}{\kappa^2}(\kappa-\mu)
    \bigg)^{1/2} +\mathcal{O}\left(\frac{1}{\beta^3}\right)
    \bigg\}\label{eq:entropy_red}
\end{align}
Let's now use differentiation by parts on the second term of Eq. \eqref{eq:entropy_red} as follows:
\begin{align}
    &\int_{m0}^\mu dE \frac{1}{3\beta}\frac{d}{dE}\left(E^2-m^2\bigg(1-\frac{E}{\kappa}\bigg)^2\right)^{3/2}(E-\mu)\nonumber\\
    &=\frac{1}{3\beta}\left(E^2-m^2\bigg(1-\frac{E}{\kappa}\bigg)^2\right)^{3/2}(E-\mu)\Bigg|_{m_0}^\mu - \int_{m0}^\mu dE \frac{1}{3\beta}\left(E^2-m^2\bigg(1-\frac{E}{\kappa}\bigg)^2\right)^{3/2}\nonumber\\
    &=- \int_{m0}^\mu dE \frac{1}{3\beta}\left(E^2-m^2\bigg(1-\frac{E}{\kappa}\bigg)^2\right)^{3/2} \label{eq:entropy_sec_red}
\end{align}
The boundary terms vanish explicitly. Inserting Eq. \eqref{eq:entropy_sec_red} in Eq. \eqref{eq:entropy_red} boils down the entropy expression to:
\begin{align}
    S=&\frac{4\pi gV}{h^3}
    k_B
    \bigg\{\frac{1}{\beta}\frac{\pi^2}{3}
    \frac{1}{\kappa^2}(\kappa m^2+\kappa^2\mu-m^2\mu)
    \bigg(
    \mu^2-\frac{m^2}{\kappa^2}(\kappa-\mu)^2
    \bigg)^{1/2} \bigg\} +\mathcal{O}\left(\frac{1}{\beta^3}\right)\label{eq:entropy_final}
\end{align}
For $0<m_0<\mu<\kappa$, the entropy is positive.
It can be seen that for exactly $T=\frac{1}{\beta}=0$, the entropy, given finally by Eq. \eqref{eq:entropy_final} vanishes to 0 as expected. Also, if we have $\kappa\to\infty$ and $T\to0$, i.e. the special relativistic case at low temperature, Eq. \eqref{eq:entropy_final} yields:
\begin{align}
    S\big|_{\rm SR}=&\frac{4\pi gV}{h^3}
    k_B \frac{1}{\beta}\frac{\pi^2}{3}
    \mu^2 +\mathcal{O}\left(\frac{1}{\beta^3}\right)\,.
\end{align}
\section{Sommerfeld expansion of internal energy}
\textbf{Internal Energy}:
\begin{align}
    U=&
    \frac{4\pi g V}{h^3}
    \int_0^\kappa \text{d}p p^2
    E
    \frac{1}{1+e^{\beta(E-\mu)}}\nonumber\\
    =&\frac{4\pi g V}{h^3}
    \int_{m_0}^\kappa \text{d}E
    \Bigg(E^2-m^2\Bigg(1-\frac{E}{\kappa}\Bigg)^2\Bigg)^{1/2}
    \Bigg(E+\frac{m^2}{\kappa}\Bigg(1-\frac{E}{\kappa}\Bigg)\Bigg)
    E
    \frac{1}{1+e^{\beta(E-\mu)}}\nonumber\\
    =&\frac{4\pi g V}{h^3}
    \Bigg\{
    \int_{m_0}^\mu \text{d}E
    \Bigg(E^2-m^2\Bigg(1-\frac{E}{\kappa}\Bigg)^2\Bigg)^{1/2}
    \Bigg(E+\frac{m^2}{\kappa}\Bigg(1-\frac{E}{\kappa}\Bigg)\Bigg)
    E\nonumber\\
    &+\frac{1}{\beta^2}
    \frac{\pi^2}{6}
    \frac{-7 \kappa m^4 \mu^2 + 3 m^4 \mu^3 - \kappa^3 (m^4 - 7 m^2 \mu^2) + 
 \kappa^4 (-2 m^2 \mu + 3 \mu^3) + 
 \kappa^2 (5 m^4 \mu - 6 m^2 \mu^3)}
{\kappa^4 \sqrt{-\left(\frac{m^2 (\kappa - \mu)^2}{\kappa^2}\right) + \mu^2}}
\Bigg\} +\mathcal{O}\left(\frac{1}{\beta^3}\right) \label{eq:U_der}
\end{align}
\section{Sommerfeld expansion of specific heat capacity}
\begin{align}
    C_V&=\left(\frac{\partial U}{\partial T}\right)_{V,\,N}\nonumber\\
    &=\frac{4\pi gV}{h^3}\Bigg\{\frac{k_B}{\beta}\frac{\pi^2}{3}\,\frac{-7\kappa m^4\mu^2+3m^4\mu^3-\kappa^3(m^4-7m^2\mu^2)+\kappa^4(-2m^2\mu+3\mu^3)+\kappa^2(5m^4\mu-6m^2\mu^3)}{\kappa^4\sqrt{-\left(\frac{m^2(\kappa-\mu)^2}{\kappa^2}\right)+\mu^2}}\nonumber\\
&-\,k_B\beta^2\,\frac{\partial\mu}{\partial\beta}\left(\mu^2-m^2\left(1-\frac\mu\kappa\right)^2\right)^{1/2}\left(\mu+\frac{m^2}\kappa\left(1-\frac\mu\kappa\right)\right)\mu \;+\;\mathcal{O}\!\left(\frac1{\beta^3}\right)\Bigg\} \label{eq:Cv_der}
\end{align}
This follows from Eq. \eqref{eq:U_der}. The term of the order of $\frac{1}{\beta^2}$ contributes to the $\frac{1}{\beta}$ order term of Eq. \eqref{eq:Cv_der}. The term of the order of $\frac{1}{\beta^2}$ of Eq. \eqref{eq:Cv_der} should come from the term proportional to $\frac{1}{\beta^3}$ of Eq. \eqref{eq:U_der} which is 0. Note that there will be another term contributing from term proportional to $\frac{1}{\beta^2}$ in Eq. \eqref{eq:U_der}. This corresponds to the derivative of $\mu$ with respect to $\beta$. But from Eq. \eqref{eq:mu_final} it can be seen that this term is of the order of $\frac{1}{\beta^3}$ and can be included in $\mathcal{O}\!\left(\frac1{\beta^3}\right)$. Now, let's define:
\begin{align}
    h(E)&=\frac{4\pi gV}{h^3}\Bigg(E^2-m^2\Bigg(1-\frac{E}{\kappa}\Bigg)^2\Bigg)^{1/2}
    \Bigg(E+\frac{m^2}{\kappa}\Bigg(1-\frac{E}{\kappa}\Bigg)\Bigg)
    E\nonumber\\
    &=\rho(E) E\,.
\end{align}
We know that Eq. \eqref{eq:U_der} can be written as:
\begin{align}
    U&=\int_{m_0}^\mu dE h(E) + \frac{1}{\beta^2} \frac{\pi^2}{6} h'(\mu) + \mathcal{O}\!\left(\frac1{\beta^3}\right)\,.
\end{align}
Hence $C_V$ in Eq. \eqref{eq:Cv_der} is nothing but:
\begin{align}
    C_V&=\left(\frac{\partial U}{\partial T}\right)_{V,\,N}\nonumber\\
    &=-k_B\beta^2 \frac{\partial\mu}{\partial\beta} h(\mu) + k_B \frac{1}{\beta} \frac{\pi^2}{3} h' (\mu) + \mathcal{O}\!\left(\frac1{\beta^3}\right)\nonumber\\
    &=-k_B\beta^2 \frac{\partial\mu}{\partial\beta} \rho(\mu) \mu + k_B \frac{1}{\beta} \frac{\pi^2}{3} \left(\rho(\mu) + \rho'(\mu) \mu \right) + \mathcal{O}\!\left(\frac1{\beta^3}\right)\,. \label{eq:Cv_param}
\end{align}
From Eq. \eqref{eq:mu} we can write the following:
\begin{align}
    \frac{\partial \mu}{\partial \beta} = 2 \frac{\rho'(\mu)}{\rho(\mu)}\frac{1}{\beta^3}\frac{\pi^2}{6}\,.
\end{align}
Using this in Eq. \eqref{eq:Cv_param}, we get:
\begin{align}
    C_V&=-k_B \frac{1}{\beta} \frac{\pi^2}{3} \rho'(\mu) \mu + k_B \frac{1}{\beta} \frac{\pi^2}{3} \left(\rho(\mu) + \rho'(\mu) \mu \right) + \mathcal{O}\!\left(\frac1{\beta^3}\right)\nonumber\\
    &= k_B \frac{1}{\beta} \frac{\pi^2}{3} \rho(\mu)   + \mathcal{O}\!\left(\frac1{\beta^3}\right)\,.\label{eq:Cv_param_final}
\end{align}
Also, from Eq. \eqref{eq:entropy_final}, we get the following:
\begin{align}
    T \bigg(\frac{\partial S}{\partial T}\bigg)_{V,N} &=\frac{4\pi gV}{h^3}
    k_B
    \bigg\{\frac{1}{\beta}\frac{\pi^2}{6}
    \frac{2}{\kappa^2}(\kappa m^2+\kappa^2\mu-m^2\mu)
    \bigg(
    \mu^2-\frac{m^2}{\kappa^2}(\kappa-\mu)
    \bigg)^{1/2} \bigg\} +\mathcal{O}\left(\frac{1}{\beta^3}\right)\nonumber\\
    &=k_B \frac{1}{\beta} \frac{\pi^2}{3} \rho(\mu)   + \mathcal{O}\!\left(\frac1{\beta^3}\right)\,.\label{eq:S_param}
\end{align}
Hence, from Eq. \eqref{eq:Cv_param_final} and Eq. \eqref{eq:S_param} it can be concluded that 
\begin{align}
    C_V=T \bigg(\frac{\partial S}{\partial T}\bigg)_{V,N}\,.
\end{align}
\section{Chemical Potential at Fixed Number Density in the High-Temperature Regime} \label{sec:appB}
In the second case, where the particle number density $n$ is fixed, the chemical potential $\mu(T, n, \kappa)$
is the unique solution to the number equation
\begin{align} \label{eq:Numberdensity}
    n & = \int_{m_{0}}^{\kappa}\, dE\, \rho(E)\, \frac{1}{e^{\beta(E-\mu)}+1},
\end{align}
where $\rho(E)$ is the MS density of states given in Eq. \eqref{eq:distri} and $\beta = \frac{1}{T}$. Since $n$ is a strictly monotonic increasing function of $\mu$ for any $T$ and $\kappa$, Eq. \eqref{eq:Numberdensity} has a unique solution $\mu(T, n, \kappa)$ for every physically admissible $n>0$. We describe its behaviour across the high-temperature regime.

For $T\gg T_{F}$, quantum degeneracy is negligible and the Fermi-Dirac distribution reduces to the Maxwell-Boltzmann form $f(E) \approx e^{-\beta(E-\mu)}$. Substituting into Eq. \eqref{eq:Numberdensity}:
\begin{align}
    n \approx e^{\frac{\mu}{T}}\, \frac{g}{2\pi^{2}}\,\int_{m_{0}}^{\kappa}\, dE\, \rho(E)\, e^{-\frac{E}{T}} \equiv e^{\frac{\mu}{T}}\, \mathcal{Z}(T,\kappa),
\end{align}
where $\mathcal{Z}(T,\kappa)$ is the single particle partition function in the MS framework. Solving explicitly, we get
\begin{align}\label{eq:1}
    \mu & = T\,\ln{\left(\frac{n}{\mathcal{Z}(T,\kappa)}\right)}.
\end{align}
Evaluating $\mathcal{Z}$ at large $T$ using $\rho(E) \approx \frac{g}{2\pi^{2}}\,E^{2}$:
\begin{align}\label{eq:2}
    \mathcal{Z}(T,\kappa) \approx \frac{g}{2\pi^{2}}\,\int_{0}^{\infty}\, dE\, e^{-\frac{E}{T}}\,E^{2} = \frac{g}{\pi^{2}}\, T^{3}.
\end{align}
Substituting Eq. \eqref{eq:2} in Eq.\eqref{eq:1}, we get
\begin{align}
    \mu \approx T\,\log{\left(\frac{\pi\,n^{2}}{g\,T^{3}}\right)} = T\,\left[\ln{\left(\frac{n\pi^{2}}{g}\right)}-3\,\ln{T}\right] + \mathcal{O}\left(\frac{T^{2}}{\kappa}\right).
\end{align}
This is the standard classical ideal gas result for the chemical potential. The dominant $ -\,3\,T\,\ln{T}$ term drives $\mu \rightarrow -\infty$ as $T \rightarrow \infty$. The MS deformation scale $\kappa$ enters through corrections of order $\frac{T}{\kappa}$ to $\rho(E)$, which contribute an additive correction of order $\frac{T^{2}}{\kappa}$ to $\mu$, but do not alter the qualitative divergence.

\bibliography{ref}

\end{document}